%% file: paper.tex
\documentclass[letterpaper,twocolumn,10pt]{article}
\usepackage{usenix,epsfig,endnotes}

\usepackage{comment}
\usepackage{multirow}
\usepackage{textcomp}
\usepackage{color}
\usepackage[normalem]{ulem}
\usepackage{setspace}
\usepackage{enumerate}

\newcommand{\cb}{\textcolor{blue}}
\newcommand{\cred}{\textcolor{red}}

\newcommand{\keyssd}{\textsc{Key-SSD}}
\newcommand{\keyftl}{\textsc{Key-FTL}}
\newcommand{\statickeyftl}{\textsc{Key-FTL(S)}}
\newcommand{\dynamickeyftl}{\textsc{Key-FTL(D)}}
\newcommand{\keylba}{\textsc{KeyLBA}}
\newcommand{\keyinode}{\textsc{KeyInode}}

\usepackage[table]{xcolor}
\usepackage{tabularx}
\newcolumntype{P}[1]{>{\centering\arraybackslash}p{#1}}
\newcolumntype{M}[1]{>{\centering\arraybackslash}m{#1}}

\newcommand{\squishlist}{
  \begin{list}{$\bullet$}
  { \setlength{\itemsep}{0pt}      \setlength{\parsep}{-0pt}
    \setlength{\topsep}{4pt}       \setlength{\partopsep}{0pt}
    \setlength{\listparindent}{-2pt}
    \setlength{\itemindent}{-5pt}
    \setlength{\leftmargin}{1em} \setlength{\labelwidth}{0em}
    \setlength{\labelsep}{0.5em} } }

\newcommand{\squishend}{
    \end{list}  }
    
    \newcommand{\squishlistenum}{
\begin{enumerate}
  { \setlength{\itemsep}{0pt}      \setlength{\parsep}{-0pt}
    \setlength{\topsep}{4pt}       \setlength{\partopsep}{0pt}
    \setlength{\listparindent}{-2pt}
    \setlength{\itemindent}{-5pt}
    \setlength{\leftmargin}{1em} \setlength{\labelwidth}{0em}
    \setlength{\labelsep}{0.5em} } }

\newcommand{\squishendenum}{
\end{enumerate} }

\begin{document}

\date{}

\title{{\keyssd}: Access-Control Drive to Protect Files from Ransomware Attacks}

\author{
{\rm Jinwoo Ahn$^1$\thanks{Both authors contributed equally to this work.} , Donggyu Park$^1$$^*$, Chang-Gyu Lee$^1$, Donghyun Min$^1$, Junghee Lee$^3$}\\
{\rm Sungyong Park$^1$, Qian Chen$^2$, Youngjae Kim$^1$\thanks{Y. Kim is a corresponding author.}}\\\\
$^1$Sogang University, Seoul, Republic of Korea, $^2$University of Texas at San Antonio, TX USA\\
$^3$Korea University, Seoul, Republic of Korea
} 


\maketitle




\input{abstract}

\vspace{-0.15in}
\input{intro}

\vspace{-0.15in}
\input{back}
\vspace{-0.3in}
\input{overview}

\vspace{-0.2in}
\input{proposed}

\vspace{-0.2in}
\input{eval}
\vspace{-0.2in}
\input{related}
\vspace{-0.4in}
\input{conc}
\vspace{-0.15in}
\input{ack}
\vspace{-0.15in}

{
\bibliographystyle{acm}
\bibliography{ref}}


\end{document}

%% file: abstract.tex
\begin{abstract}
Traditional techniques 
to prevent
damage from ransomware attacks are to detect and block attacks by 
monitoring the known behaviors such
as {frequent} name changes, {recurring} access to cryptographic libraries
and exchange keys with remote servers. 
Unfortunately, intelligent ransomware can easily bypass these techniques. 
Another prevention technique is to recover from the backup copy when a file is infected with ransomware. 
However, the data backup technique requires extra storage space and can be removed with ransomware.
In this paper, we propose to implement an access control mechanism on a disk drive, called a {\keyssd} disk drive. 
{\keyssd} is the data store and the last barrier to data protection.
Unauthorized applications will not be able to read file data even if they bypass the file system defense, thus denying the block request without knowing the disk's registered block key and completely eliminating the possibility of the file becoming hostage to ransomware.
We have prototyped {\keyssd} and
validated the usefulness of {\keyssd} by demonstrating (i) selective block access control, (ii) unauthorized data access blocking and (iii) negligible performance overhead. 
{Our comprehensive evaluation} of {\keyssd}  for various workloads show the {\keyssd} performance is hardly degraded due to OS lightweight key transmission and access control drive optimization.
We also confirmed that {\keyssd} successfully protects the files in the actual ransomware sample.
\end{abstract}

%% file: intro.tex
\section{Introduction}
\label{sec:intro}
\vspace{-0.05in}

Ransomware is a type of malware that encrypts data files of a victim computer and 
requires the victim to pay a ransom to regain file access. 
Very recently, in June 2017, more than 12,000 computers were attacked worldwide by ransomware including those of at least 80 large companies~\cite{a_new_ransomware_outbreak}. 
Only a month before, a similar massive ransomware attack happened~\cite{a_new_ransomware_outbreak}. 
According to a recent report~\cite{meeting_the_threat_of,roundtable_ransomware}, it is estimated that \$200 million were paid for ransomware only in the first quarter of 2016. 
Since ransomware incurs immediate financial damages, it is one of {the} growing concerns in information security.
A typical ransomware attack reads and encrypts files and takes encrypted files as hostages.
These attacks can be performed through the normal file I/O path of the OS or by bypassing the OS directly in the user application.

The techniques to detect and prevent ransomware has been researched and developed, but it can not be perfect as ransomware evolves.
Ransomware can be detected by monitoring the behavior of applications (potential ransomware) in the operating system, network, or file system~\cite{7536529,7600214,7336353,7784627,7736455,7764294,7387902,Continella:2016:SSR:2991079.2991110}. 
For example, if an application exhibits frequent renaming, frequent access to cryptographic library, and communicating with known malicious servers, it is considered as ransomware.
However, if adversaries are aware of these techniques, they may manage to develop a new type of ransomware that does not exhibit the behaviors recognized by these techniques. 
Another way to mitigate ransomware is to back up data. 
If a file is infected by ransomware,
version control systems 
can be
used to track the history of a file and recover it. 
However, maintaining backup copies of files requires additional storage and may incur the performance overhead on host computers and network traffic. 
Furthermore, there is a risk of intelligent ransomware destroying backup files~\cite{7924925,7886569}.

To address the problems of these existing techniques, we present a fundamental solution of ransomware that is not based on signature-based behavioral monitoring (which can be circumvented by intelligent ransomware) and does not cause excessive storage or performance overhead.
In this paper, we present {\keyssd}, an {\em access-control drive} to protect the files from ransomware attacks.
{\keyssd} can shield 
aforementioned attacks through I/O paths to the disk drive. 
Our primary contribution is that {\keyssd} implements a {\em disk-level access-control mechanism} where unauthorized requests are denied by a \textit{disk drive}. 
Even if ransomware bypasses the traditional access-control mechanism in the file system by exploiting the vulnerabilities of applications, it cannot bypass the disk-level access-control mechanism.
Unauthorized ransomware cannot even read 
a file data on the disk drive, and consequently, cannot take files hostage.
Traditional SSDs are block-devices that do not have capability of granting or denying read/write requests.
On the other hand, {\keyssd}
controls block-level access to block-level read/write requests, thereby blocking unauthorized data read/write access.
Object storage~\cite{1612479} includes an \textit{object-level} access-control mechanism.
Since it adopts a higher level of abstraction, 
it requires a significant modification to the system software to take advantage of it.
But, {\keyssd} 
requires only minimal modification of existing system software because it maintains the traditional abstraction level.

While traditional access-control mechanisms are implemented in a \textit{file system} and control access to \textit{files}, our focus is on controlling access to \textit{blocks} by a \textit{disk drive}.
Modern SSDs use low-power and
multi-core 
controllers to provide significant computing performance~\cite{tiwari_active_2013, cho2013, 
Gu:2016:BFN:3007787.3001154, Qin:2006, boboila2012active, sim:2015:sc}. 
For {\keyssd}, we implemented the access control mechanism using 
these powerful computation resources
on the disk drive. 
Specifically, 
the disk-level access control mechanism is implemented in the flash translation layer (FTL) on the SSD. 
Blocks in any file that need protection from Ransomware attacks can be assigned an access code (key) from the application. The FTL of an SSD determines whether to grant access to each block while retaining key information per block. 
Since the key must be delivered to disk through the OS kernel, it is necessary to modify the OS, but it is crucial to implement it with little performance degradation.

{We conducted a comprehensive evaluation 
for {\keyssd}
by implementing the LBA-key map table in Linux kernel while keeping the existing SATA protocols with real commodity SSD and the Jasmine OpenSSD platform~\cite{jasmine}. 
We compared the performance of 
{\keyssd}
with a variety of file I/O patterns using a mix of synthetic and realistic workloads. 
Specifically, in our evaluation with both real SSD and Jasmine OpenSSD platform, we observed that {\keyssd} yields negligible overhead compared to a baseline with normal SSDs without the access-control mechanism.
We have also verified that {\keyssd} can block unauthorized ransomware I/O access by running actual ransomware code.}

%% file: back.tex
\section{Background}
\label{sec:back}

\vspace{-0.1in}
\subsection{Threat Model: Ransomware Attacks}
\label{sec:threat}
\vspace{-0.05in}

The primary target of ransomware is data files created by a user through an application (e.g. Word processor). 
As will be discussed in more detail in Section~\ref{sec:sec_anal}, typical ransomware goes through infection, persistence, removing backup copies, encryption, and notice.
Specifically, we consider that ransomware may exhibit following behaviors.

\squishlist
\item While encrypting a user file, ransomware may overwrite it or create a new file after deleting the original file.
Though file deletion only removes the metadata that keeps data blocks in the disk, the data blocks may be overwritten by other write operations.
Thus, we consider both cases to ensure no data loss.
\item Ransomware may access files through a regular file system or directly access the raw disk drive without going through the file system.
Traditional access-control mechanisms are implemented in the file system.
If ransomware bypasses the file system, there is no more barrier in the disk drive.
Since the proposed access-control mechanism is implemented in the disk drive, it can also prevent direct access attacks.
\item Ransomware may be a user-level application with a root privilege.
Ransomware can acquire a root privilege by exploiting vulnerabilities in applications or operating system.
Once it has a root privilege, it can access files of other users.
The proposed technique can defend files against this type of ransomware.
\item Ransomware may hijack system calls, but cannot access kernel data structures in file system layers.
System call hijacking is one of the most popular techniques to implement a rootkit (kernel-level malware).
A system call may be replaced by a malicious one.
However, it is assumed that kernel data structures 
cannot be tampered because it requires recompilation of the kernel, which is more challenging than hijacking.
\squishend


\vspace{-0.15in}
\subsection{Access Control in Disk Drive}
\label{sec:justification}
\vspace{-0.05in}

Data files created by a user cannot be restored unless they have backup copies. 
Loss of data incurs not only financial damages but also interruption of operations.
In 2015, a zero-day ransomware, WannaCry, attacked computers in more than 150 countries, and caused U.K. National Health Service hospitals and Honda Motor Company to shut down~\cite{chen_robert}.
This catastrophic damage implies that existing countermeasures were not effective.
The application may protect data files by a password or encryption, but a password and encryption are intended to protect \textit{contents} of a file from being revealed, not a file itself. 
Thus, ransomware is still able to read and encrypt the file, though the ransomware may not interpret the contents of the file. 

The proposed approach is to integrate an access-control mechanism with a disk drive.
The access-control mechanism allows access to files under protection only for authorized applications.
This access-control mechanism specifically targets at preventing ransomware.
In fact, this access-control mechanism can be implemented in the file system.
But, 
Compared to file system implementation, disk-level implementation has the following advantages.

\textbf{Security:}
Since a disk drive is a separate system from a host, it is not easy to compromise both host and disk drive at the same time.
Especially, if the disk drive does not allow firmware update by the host, it is very difficult to compromise the disk drive unless the disk drive is physically accessible.

\textbf{Compatibility:}
The implementation of the disk-level access-control mechanism is independent of a file system.
To support {\keyssd}, an additional kernel module needs to be inserted to the file system, but it does not change any structures in the file system.
Therefore, {\keyssd} can be adopted without changing the file system.

%% file: overview.tex
\section{{{\keyssd}}: Access Control Drive}
\label{sec:overview}
\vspace{-0.1in}

\subsection{Goals}
\vspace{-0.1in}

In this section, we discuss 
our key design principles. 

{\bf Selective Disk-level Block Access Control:}
Data blocks must be given an access key to grant access to the data block. This access control must be implemented within the disk drive. 
%
Not all data blocks need to be protected with an access key. The user/application should be able to specify data blocks of the files to be protected, and the data blocks must be selectively protected by the access key according to the user's request. 
It is also necessary to minimize the performance and space overhead of managing the key per block in the disk drive. 
In addition, the cost of keeping this information persistent should be minimized.

{\bf Key Transmission in OS:}
Since the disk drive controls the block-level key access, the OS must be able to transfer the key corresponding to the block to the disk drive.
Also, data blocks of the protected files and their corresponding keys must be transferred to the disk drive at the same time. 
There is a possibility for ransomware to access the disk without a key if data block and key transfer occur separately. 

{\bf Minimal OS support for {\keyssd}:} 
Ransomware can attack through normal file I/O paths in OS or bypassing the OS file system to access file data directly.
In order to differ normal file I/O requests and direct I/O requests, the OS must be able to assign the keys only to the block requests of normal file I/O operations. 
In addition, most OS speed up disk access through the file system page cache. Because the page cache does not require disk access, malicious attacks can not only read data using cached data, but also delete files without accessing disk data.
To protect against attacks using data in the page cache, an OS level implementation is required.
To this end, OS kernel code modifications should be minimized.


\vspace{-0.1in}
\subsection{{{\keyssd}} Overview}
\vspace{-0.05in}

\begin{figure}[t!]
\centering
\includegraphics[width=0.33\textwidth, trim={0.5cm, 0.5cm, 0.5cm, 0.5cm}]{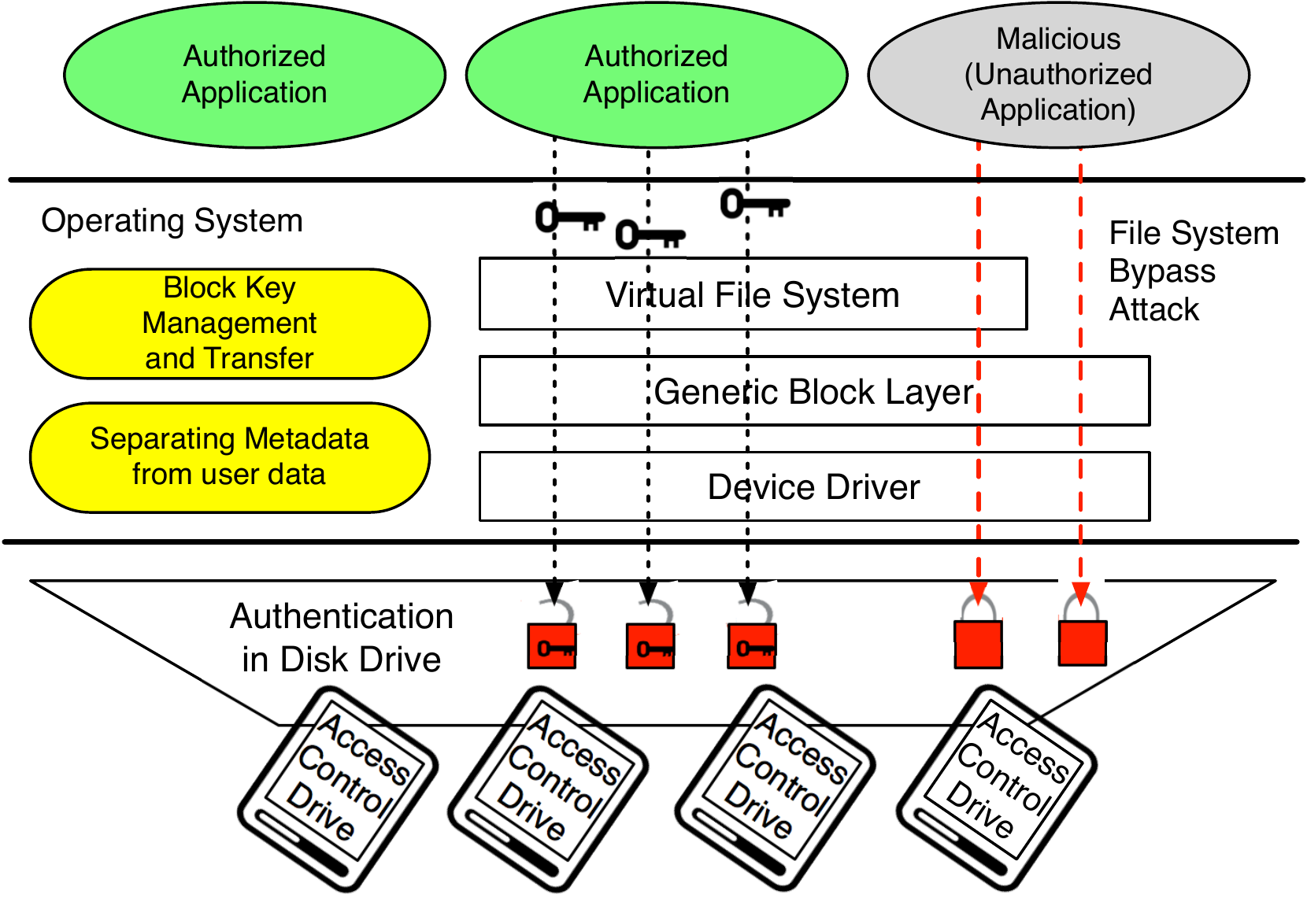}
\vspace{-0.05in}
\caption{An overview for {\keyssd} with OS stack. 
}
\label{fig:overview}
\vspace{-0.15in}
\end{figure}

We envision {\keyssd} to be the last firewall to block ransomware attacks.
Figure~\ref{fig:overview} presents a bottom-up description of the system for each component necessarily implemented at every level of OS stack and device drive. 

{\bf Access Control Drive:} The lowest level is the access-control drive ({\keyssd}) that is a solid-state drive, capable of running {disk}-level access-control. Flash translation layer (FTL), which is a firmware in the SSD, manages the key per block.
To authorize a request, an application needs to put a request along with a key.
If a key is sent 
separately from the request, the key can be exploited by ransomware by putting a malicious request between the key and the valid request.
Therefore, we guarantee a request arrives at the disk drive at the same time with a key by piggy-backing the key with the request in the SATA protocol.

{\bf OS Support for Access Control by {\keyssd}:}
OS manages the key and passes the key assigned by the application to the access control drive. 
The key stays in the OS temporarily. 
In particular, when a block request is transmitted to {\keyssd} in a block layer, the type of block (normal file I/O operation or direct I/O access)
can be distinguished and a key can be assigned accordingly. 
Moreover, it also protects against attacks using the OS's page cache and deleting files without authentication by {\keyssd}.

{\bf Application Interface Passing Keys to OS:}
The user must be able to assign keys to blocks of files to be protected. Especially, applications need a mechanism to transfer block keys to OS. In other words, it should not go beyond the existing system call interface design principles and it should be able to pass the file block keys to the OS when performing I/Os. 

{\em Together, these construct to build a last-level protection against data attacks by malicious applications.}

%% file: proposed.tex
\section{Design and Implementation}
\label{sec:proposed}
\vspace{-0.1in}

In this section, we describe 
the implementation of the access control mechanism in the SSD and the implementation in the OS kernel to pass keys from the application to the {\keyssd}.
Specifically, we have implemented a LBA-key management framework in OS kernel and the FTL extension for {disk}-level access control management on the SSD with the following main design goals:
(i) efficient implementation of FTL performing {disk}-level key management and access control on the SSD,
(ii) lightweight key management in OS kernel, and (iii) no modification of existing SATA protocol to communicate between host OS and the SSD.

\input{acd}
\input{key_transfer}

\input{device_comm}



\vspace{-0.15in}
\subsection{OS Security Issues}
\label{sec:issue}

\textbf{Inode protection:}
In {\keyssd}, only the user data of the file is protected by the key, and the metadata of the file is not protected. Therefore, the file open which reads only the \texttt{inode} of the file can succeed without the file key.
To solve this problem, we propose a read-verify method. The read-verify method generates a request to read data directly from the disk and confirms whether the data can be actually read. That is, it does not read only the \texttt{inode} of the file when the file is opened, but verifies that the actual data of the file can be read by using the read-verify method.

\textbf{Page cache:}
Most OS implement disk cache, called page cache, to improve disk access time. However, page cache can cause security problems to access files cached in the page cache without a key.
Consider a case where a process with the correct key reads a file, and another process without an valid key or key accesses the file from the page cache. 
In particular, a ransomware attack can read a file's original data from the page cache, encrypt it, create a new file, and delete the original file.
To solve this problem, we have implemented the read-verify method described as a solution to the \texttt{inode} protection security problem.
Previously, the read system call first checks whether there is data in the page cache of the file, and then reads the data from the disk only if it does not exist. On the other hand, the read-verify method unconditionally reads the first page data of the file from the disk, and verifies that the data can be actually read from {\keyssd} when reading the file.

\textbf{File deletion with invalid key:}
When deleting a file, we used the read-verify method by calling a function that checks if the data can be read from {\keyssd}. It can not delete a file without a valid key.

%% file: acd.tex
\vspace{-0.15in}
\subsection{Disk-level Access-Control}
\label{sec:auth_disk}

{\keyssd} 
implements a disk-level access-control mechanism where unauthorized requests are denied by the disk drive.
Access authentication is performed using 
such computational resources as an ARM-based, multi-core storage controller on SSDs. 
SSD communicates with a host through various I/O protocols such as SATA protocol. 
In this paper, we implemented {\keyssd} for SATA-based SSD.

\begin{figure}[!t]
	\begin{center}
		\begin{tabular}{@{}c@{}c@{}}
			\includegraphics[width=0.45\textwidth]{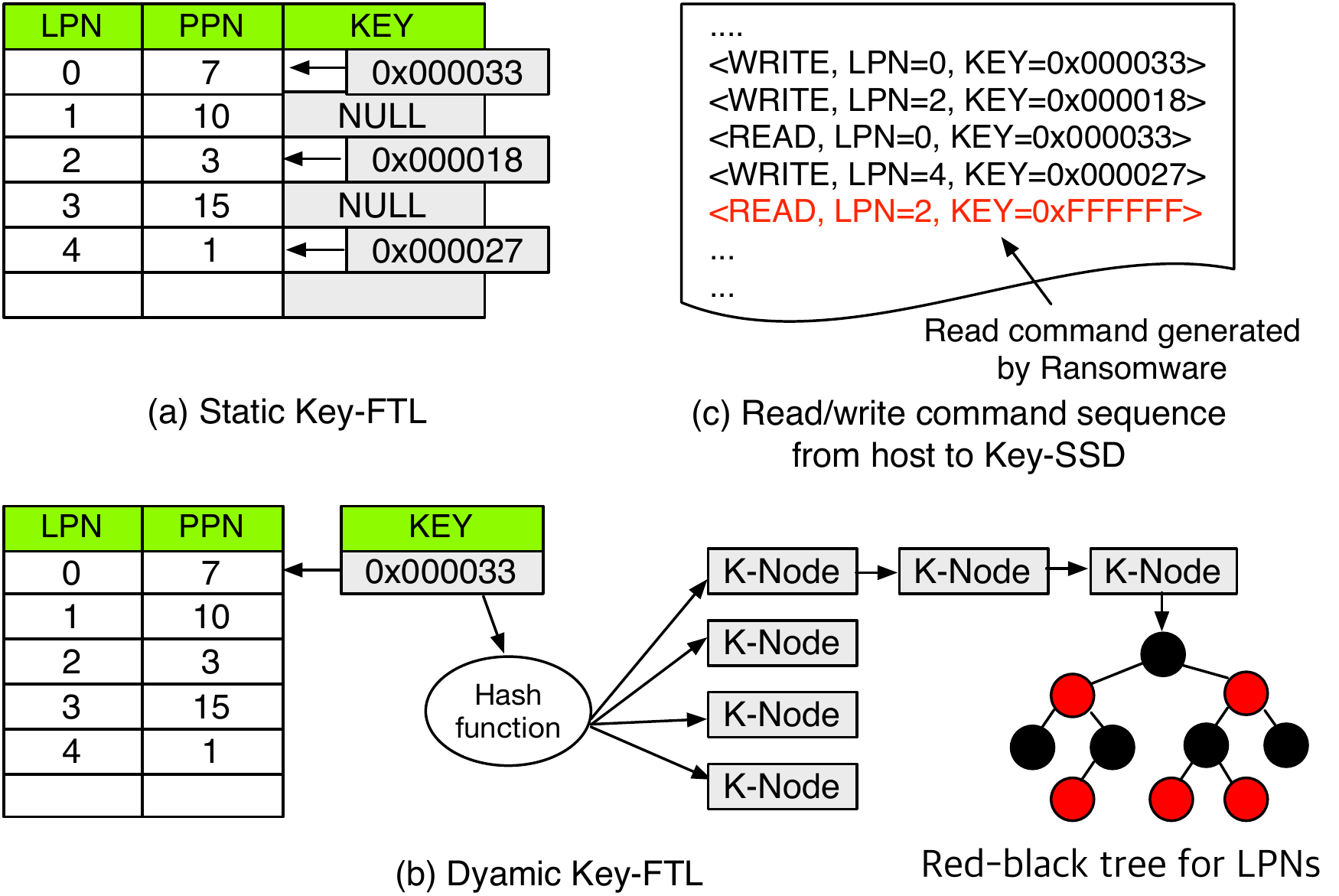} & 
\vspace{-0.1in}
		\end{tabular}
		\caption{
		(a) and (b) depict{\statickeyftl} and {\dynamickeyftl}. In (b), K-Node is a pointer to a red-black-tree for LPNs corresponding to a key, and a node in the red-black-tree denotes a LPN.
		(c) shows sequence of read/write operations with access-codes. 
		}
		\label{fig:key-ftl}
		\vspace{-0.3in}
	\end{center}
\end{figure}

{\bf {\keyftl}:}
{\keyssd} can extend the FTL to manage the key per block, for which we call {{\em \keyftl}}.
{\keyftl} can be implemented statically or dynamically, depending on how the key per block is managed in the SSD's internal memory.
In a static method, a key field can be added to each mapping table entry to have a unique key for each {disk} block. 
This method, called {\statickeyftl} is advantageous in that the implementation is simple and the key search time for block access is O(1) because FTL is a linear page-table, as shown in Figure~\ref{fig:key-ftl}(a).
However, 
the drawback is that a large memory space is required
because blocks of files that do not need to be protected with a key also require memory space for the key value.
In order to reduce memory space overhead, we propose a dynamic method of managing only LPNs protected by keys in the memory of an SSD ({\dynamickeyftl}).
In our implementation, this method used 
a red-black tree for managing LPNs that is protected by the corresponding key because the red-black tree data structure is appropriate for fast searching of LPNs corresponding to keys.
When a block request arrives with LPNs and key, 
the key value of the request is hashed to search for the red-black tree of the corresponding LPNs.
The search for LPNs corresponding to the key is performed.

{\bf Access Control Mechanism:}
A disk drive that receives a key along with an I/O request from the host performs access control on the data block by comparing its key value in the {\keyftl} with the key value in the requested block when performing a read/write operation.

When the device receives a write/read command, the authentication is granted as follows.
In case of a write, 
when an SSD receives a write request, 
the SSD firmware first reads the key from the {\keyftl} and compares it with the key sent by the host. 
If the value of a key is different, it sends an error message back to host without performing the write operation because it is not authenticated. 
If the key value is the same as that in the {\keyftl}, write operation is allowed. 
Also, the value of the key field of the {\keyftl} may be NULL. In this case, it means that write to the LBA has not yet occurred. Therefore, in this case, the key value received from the host is added to the key field and write is performed.
Read also operates the same way as write. 
But, in read, if the key field of the LBA referenced in the {\keyftl} is NULL, 
the key is not stored and is granted. 
Figure~\ref{fig:key-ftl}(c) shows the read write sequence with keys and how a read request generated by ransomware is denied. Refer to $<$READ, LPN=2, KEY=0xFFFFFF$>$ in the sequence. The key value of LPN=2 is 0x000018 in {\statickeyftl} in (a). 

In particular, the way to access {\dynamickeyftl} is divided into insert and search steps. Insert step is to add a new (key, LPN) to the {\dynamickeyftl}. When the host attempts to write to the new page on the SSD with the key, the corresponding LPN and key are inserted into the {\dynamickeyftl}. Search step is used for authentication by searching the {\dynamickeyftl} with (key, LPN). In our implementation, in order to minimize the overhead of sequential I/Os for authenticating every access to LPNs, we only allowed the first LPN access during search step. 

{\bf Selective Flush for {\keyftl}:}
{\keyftl} is loaded into volatile memory in SSD. Sudden power-failures can cause all {\keyftl} entries in the memory to be lost, so they need to be synchronized with flash, the permanent storage space. This operation is called {\em flush} operation. 
The {\keyftl} flush function is called from the SSD firmware. 
For the SATA protocol, HOST sends the command ATA\_CMD\_FLUSH 
and the SSD firmware calls the {\keyftl} flush 
function.
Calling a flush operation can affect SSD performance and the performance overhead is proportional to the size of the mapping table. 
The {\keyftl} additionally manages the key information 
and the size of the {\keyftl} is larger than normal FTL. 
In order to minimize the increased flushing overhead due to large FTL, we 
propose a method, called {\em Selective Flush} which flushes only the changed entries in {\keyftl}, rather than updating all the {\keyftl} entries in batches. This can greatly reduce the overhead of writing FTL to the flash. 

%% file: key_transfer.tex
\vspace{-0.15in}
\subsection{Block Key Transfer and OS Support}

To perform key authentication per block request on a disk drive, it is essential to transfer the key to the disk drive. 
The key corresponding to the LBA requested by the host must be correctly registered in the {\keyftl} of the \keyssd, and the block access authentication control must be properly performed. 
Importantly, the key must be sent synchronously with the LBA block request. 
After the block request is completed, the key must be deleted from the OS kernel to minimize space overhead. 

{\bf Key Management in OS:}
Specifically, the LBA key must pass through OS kernel including the file system, generic block layer, and block device driver.
LBA keys must be managed by the OS kernel.
We create and manage two hash tables -- KeyInode table ({\keyinode}) and KeyLBA table ({\keylba}) in the kernel for this  purpose.
In the Linux kernel, the information accessed at each layer is limited. That is, the LBA information of the file is not known at the VFS 
layer, but the file \texttt{inode} information of the VFS layer can be accessed at the generic block layer. 
Therefore, in this study, we implement the two tables in the kernel.
In Linux, each file is represented by an \texttt{inode}. 
The {\keyinode} table is implemented in the VFS layer of the kernel and manages the key value assigned to each file.
The {\keylba} manages the key value assigned to each LBA and is implemented at the generic block layer in the kernel. 
The {\keyinode} is referenced by the LBA's \texttt{inode} at the generic block layer, and it builds the {\keylba}. 
When making a request to the device from the driver, it consults with the {\keylba} to find the appropriate key to the LBA and send the LBA and key together at the same time in a request to the device.

\begin{figure}
\centering
\includegraphics[width=0.4\textwidth, trim={0.5cm, 0.5cm, 0.5cm, 0.5cm}]{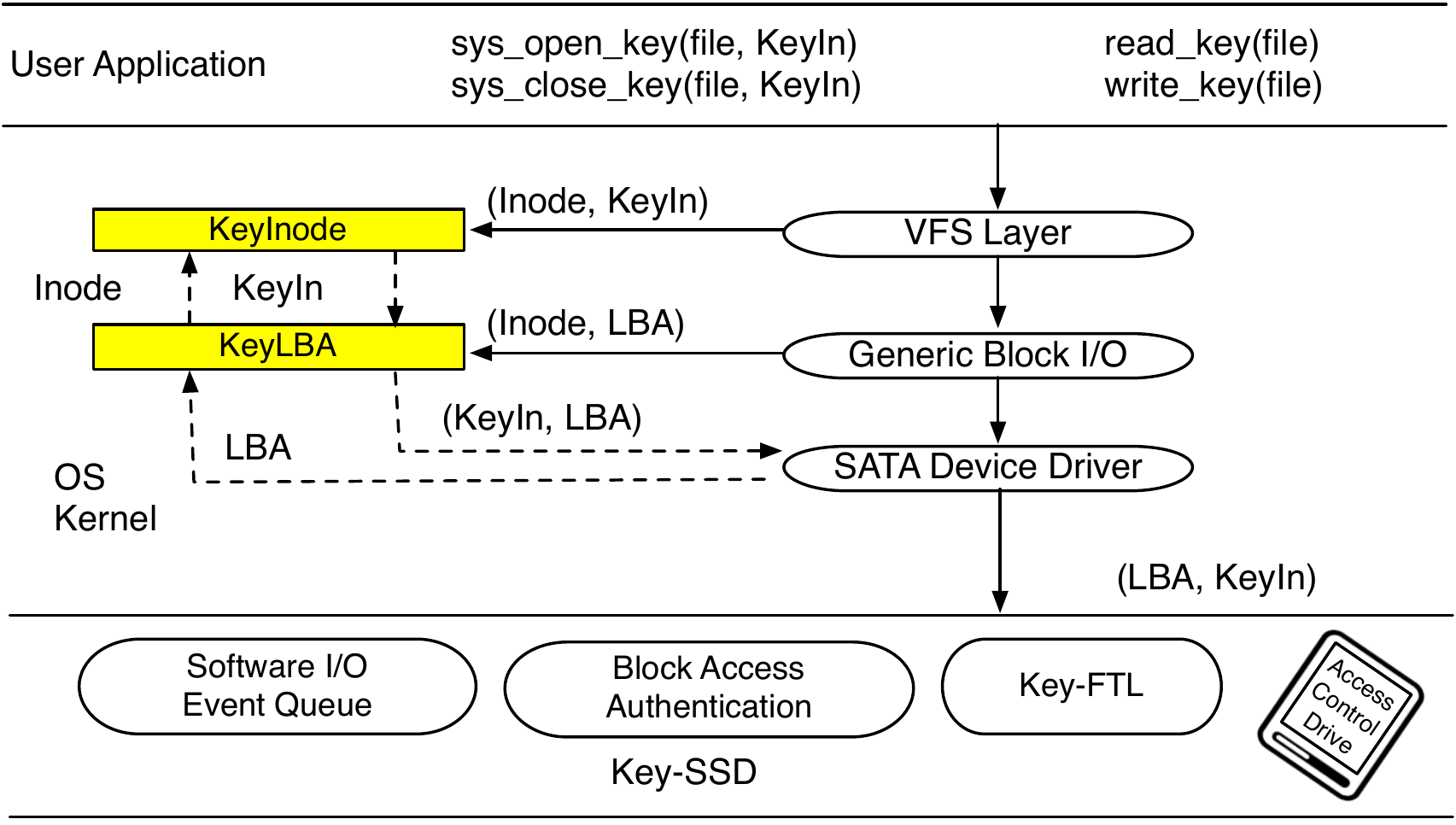}
\vspace{-0.05in}
\caption{
Illustration of block key transfer with 
{\keyinode} and {\keylba}
in OS kernel.
\texttt{KeyIn} is an input key of a file specified by users. 
}
\vspace{-0.15in}
\label{fig:key_transfer}
\end{figure}


An application has key information for each file. 
In order to transfer the key of the file from the application to the disk drive, we define the following system calls:  
\texttt{sys\_open\_key()}, and \texttt{sys\_close\_key()}.
Each system call is similar to existing system calls, \texttt{sys\_open()} and \texttt{sys\_close()}. 
The \texttt{sys\_open\_key()} system call can insert a new key into the {\keyinode} by sending a key from the application to the kernel when opening the file.
The \texttt{sys\_close\_key()} system call removes dynamically allocated inode-key elements from {\keyinode}
when the file is closed.
This makes the kernel free from the threat of hackers because it does not have the key information of the closed file.
Our specific implementation is as follows:
We extended \texttt{do\_sys\_open} kernel function 
by adding a key function parameter and name it \texttt{do\_sys\_open\_key}.
The \texttt{do\_sys\_open\_key} function is the first kernel function to be called when the \texttt{sys\_open\_key} system call is called. 
It uses file path to fetch file metadata such as \texttt{inode} from disk, link it to file descriptor, return it, and open file. 
In the \texttt{do\_sys\_open\_key} function,
we can create an element with key and \texttt{inode} values for the file and insert it to the 
{\keyinode}.
We also implemented \texttt{\_\_close\_fd\_key} to free the element corresponding to the current file from the 
{\keyinode}. The \texttt{\_\_close\_fd\_key} function is the first kernel function that is called when the \texttt{sys\_close\_key} system call is called, and performs file close by freeing file metadata mapped to the file descriptor. Note that the file key information is removed from the kernel by deleting the element from the {\keyinode} when the file closes. 

The generic block layer creates a 
{\keylba}
to transfer the keys to the device driver. 
Each element in the 
{\keylba}
is created when reading or writing a block of related files and is deleted after 
the key is sent from the device driver (SATA driver) to the disk drive. 
Our specific implementation is as follows:
In the generic block layer, the 
{\keylba}
is constructed using the \texttt{make\_generic\_request} function. The \texttt{make\_generic\_request} function is used to create an I/O request using the \texttt{bio} structure, which is called from the generic block layer after the EXT4 file system.
The \texttt{bio} structure has the inode information of the corresponding block.
Therefore, it is suitable to create 
{\keylba}
in this layer, and it can refer to the existing 
{\keyinode}
in this function, find the LBA key, and insert the element with LBA and key information together into the 
{\keylba}.
As we will describe in the next section, the elements of the 
{\keylba}
are deleted from the table after the device driver completes the I/O request to the device.
Figure~\ref{fig:key_transfer} illustrates the implementation of {\keyinode} and {\keylba} in the OS kernel stack.

The {\keylba} managed by the kernel can protect only the user data of the file. That is, 
the key value of {\keylba} is to protect only the data block defined by the users.
Therefore, in this study, the block type 
(data blocks requested through normal file I/O operations and those requested by directly I/O operations)
are 
distinguished in the block layer.
At the block layer in Linux, the block accesses 
are managed by \texttt{address\_space} structure of \texttt{inode}. 
If the block is mapped to the address space of the \texttt{inode} of its corresponding file, it means block requests from normal file I/O operations. 
Keys are assigned accordingly by consulting with {\keylba}.
Else if the block is mapped to the address space of the unique \texttt{inode} within the block device structure, 
it is block requests by direct I/O operations, thus keys are not assigned to those blocks. 

Secure key management is orthogonal to the proposed approach.
In this paper, we assume the key of user data is managed by the application.
To enhance key security, we may consider using a remote authentication server or Trusted Computing Module (TPM) to manage keys for {\keyssd}.



%% file: device_comm.tex
\begin{figure}
\centering
\includegraphics[width=0.48\textwidth, trim={0.5cm, 0.5cm, 0.5cm, 0.5cm}]{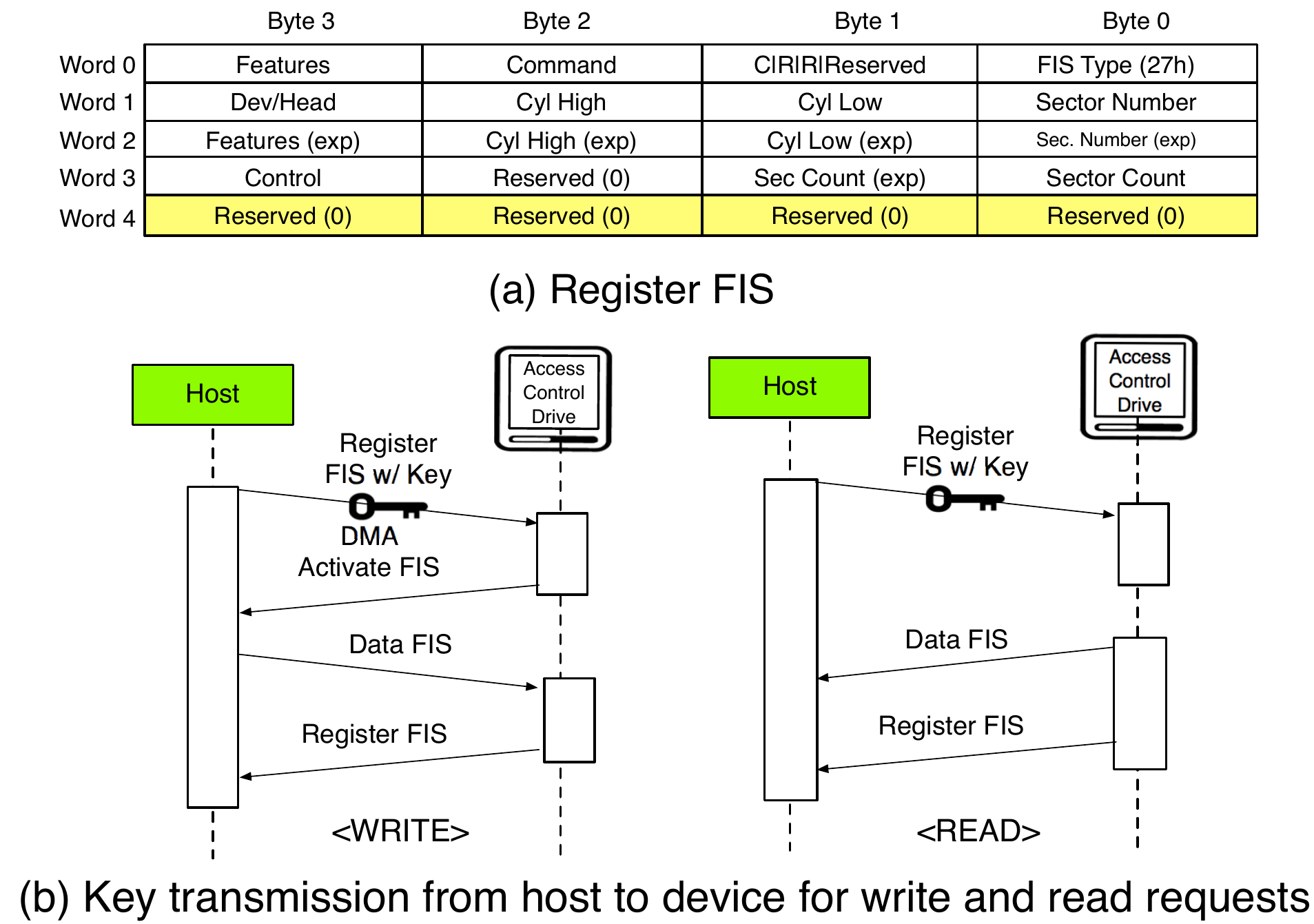}
\vspace{-0.2in}
\caption{
Illustration of key transmission between host and \keyssd~\cite{sata_book}. Word 4 in the Register FIS is used to deliver the key.
}
\label{fig:key_device_transmission}
\vspace{-0.2in}
\end{figure}

\vspace{-0.15in}
\subsection{Host to Device Communication}

The key transferred from the kernel to the SATA device driver is transmitted to the disk drive in compliance with the SATA protocol.
SATA protocol consists of 5 layers (Application, Command, Transport, Link, Physical layers). 
In the Application layer, the host stores the disk command in the shadow command register. Next, the Command layer uses the Command Sequence State Machine to create a frame information structure (FIS) transfer protocol for transmitting commands of the host.
There are a total of 14 FIS packets in the transport layer and the FIS used for disk read/write includes Register FIS and Data FIS.
In the Transport layer, the FIS is transmitted in the order specified by the Command layer, and the corresponding FIS is encoded in the Link layer and reaches the disk drive through the physical layer. In our implementation, 
the key is transferred to the disk using the transport layer of the SATA protocol.
In particular, we use the reserved space of the Register FIS to transmit the key. 

Figure~\ref{fig:key_device_transmission}(a) shows the structure of Register FIS. 
It consists of five words, but the last word (32 bit) is not used. 
Since the Register FIS is transmitted to disk drive firstly including LBA before data transfer, 
Register FIS can transmit the key synchronously with LBA and can make the key mapping table on the {\keyssd}  before data transfer. 
In addition, since the key is allocated to the reserved space of the Register FIS, there is no need to modify the SATA protocol itself for key transmission.
32-bit access can be vulnerable to brute force attacks (exhaustive search). However, {\keyssd} can implement a mechanism that counts the number of invalid attempts and rejects these brute force attacks by blocking requests if the request exceeds a predefined threshold.
And if the SATA protocol is extended, a longer key can be used.
As illustrated in Figure~\ref{fig:key_device_transmission}(b), the Register FIS is involved at the beginning and end of the protocol in both write and read operations and a key is piggybacked in the Register FIS from the host to \keyssd.

%% file: eval.tex
\section{Evaluation}
\label{sec:eval}
\vspace{-0.1in}

In this section, first we show the overhead analysis of {\keyftl} and the effectiveness of selective flushing of the {\keyftl}.
Second, we analyze the OS kernel overhead for key transmission to {\keyssd} and the end-to-end overhead including the OS kernel and {\keyftl} implementation overhead. 
Third,  we perform a step-by-step analysis of the usefulness of the {\keyssd} for a typical ransomware attack.

\vspace{-0.15in}
\subsection{Experimental Setup}
\label{sec:expr_setup}
\vspace{-0.05in}

{\bf Implementation:} 
In order to prototype the {\keyssd}, we modified 360 lines of C code in the Linux VFS and the generic block I/O layers, 180 lines of C code in the firmware of the Jasmine OpenSSD platform~\cite{jasmine}. 
We also modified 70 lines of C code in the SATA device driver. 
Specifically, {\keyftl} has extended a page-based FTL with greedy GC on 
the Jasmine OpenSSD platform. 

\begin{figure}[!t]
\centering
\begin{tabular}{@{}c@{}}
\includegraphics[width=0.38\textwidth, trim={0.5cm, 0.5cm, 0.5cm, 0.5cm}]{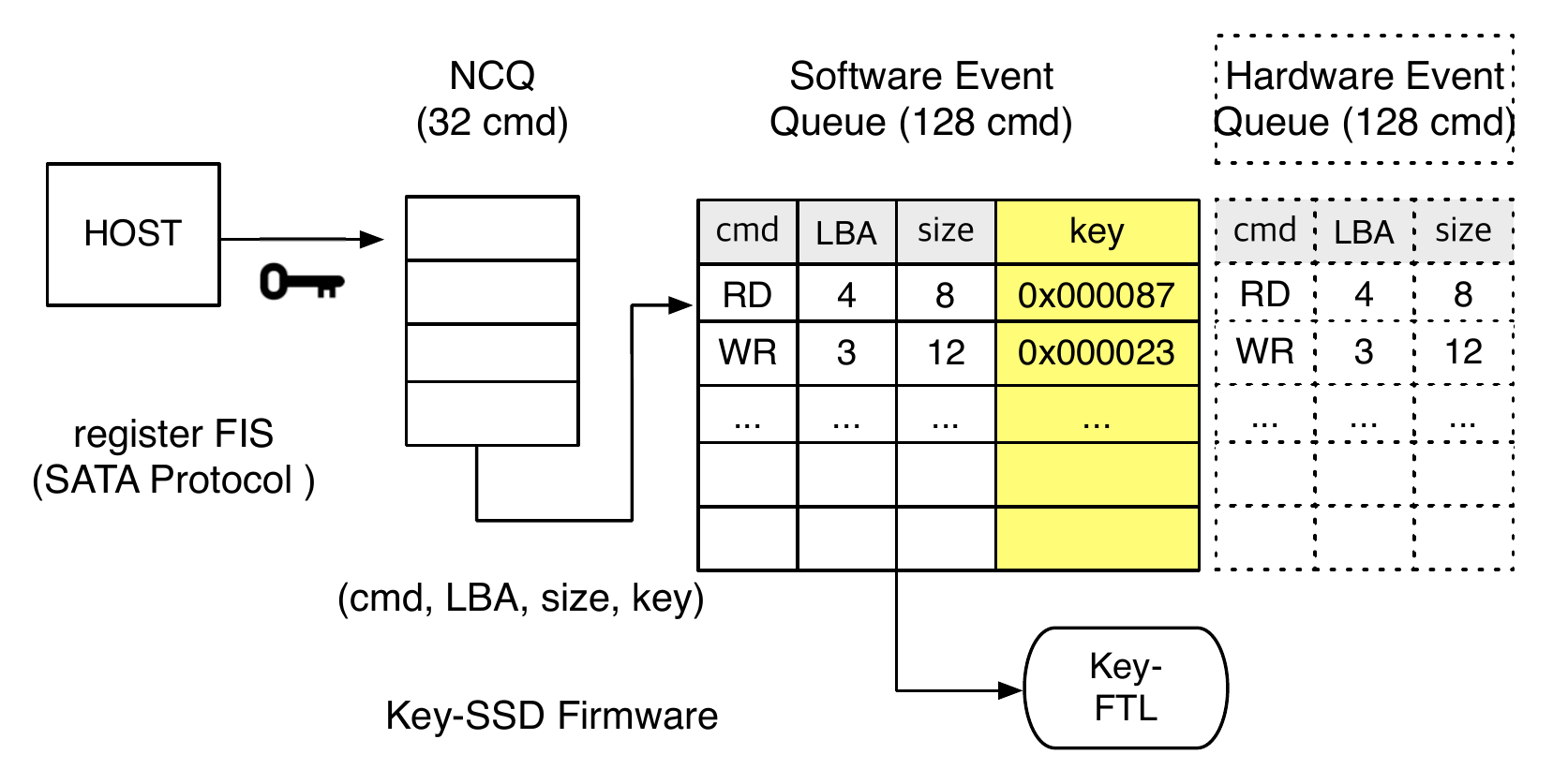}
\end{tabular}
\vspace{-0.05in}
\caption{
Software queue implementation for {\keyssd}.}
\vspace{-0.2in}
\label{fig:software_queue}
\end{figure}

{\bf Software I/O Event Queue:}
In the Jasmine OpenSSD platform, 
the command in the Register FIS sent through the SATA protocol is first assigned 
to the I/O event queue, 
which is responsible for queuing commands.
This event queue can accommodate up to 128 
I/O commands simultaneously, 
operating in FIFO mode. 
Jasmine OpenSSD platform uses a hardware event queue. 
In the hardware event queue, only the command, LBA, and size are stored automatically in hardware, so there is no way to store the key here. 
Therefore, we created a new key event queue that operates with software and stored command, LBA, size and key in the queue. 
Figure~\ref{fig:software_queue} illustrates the implementation of software event queue for {\keyssd}.

\begin{figure*}[!t]
	\begin{center}
		\begin{tabular}{@{}c@{}c@{}c@{}c@{}}
			\includegraphics[width=0.25\textwidth]{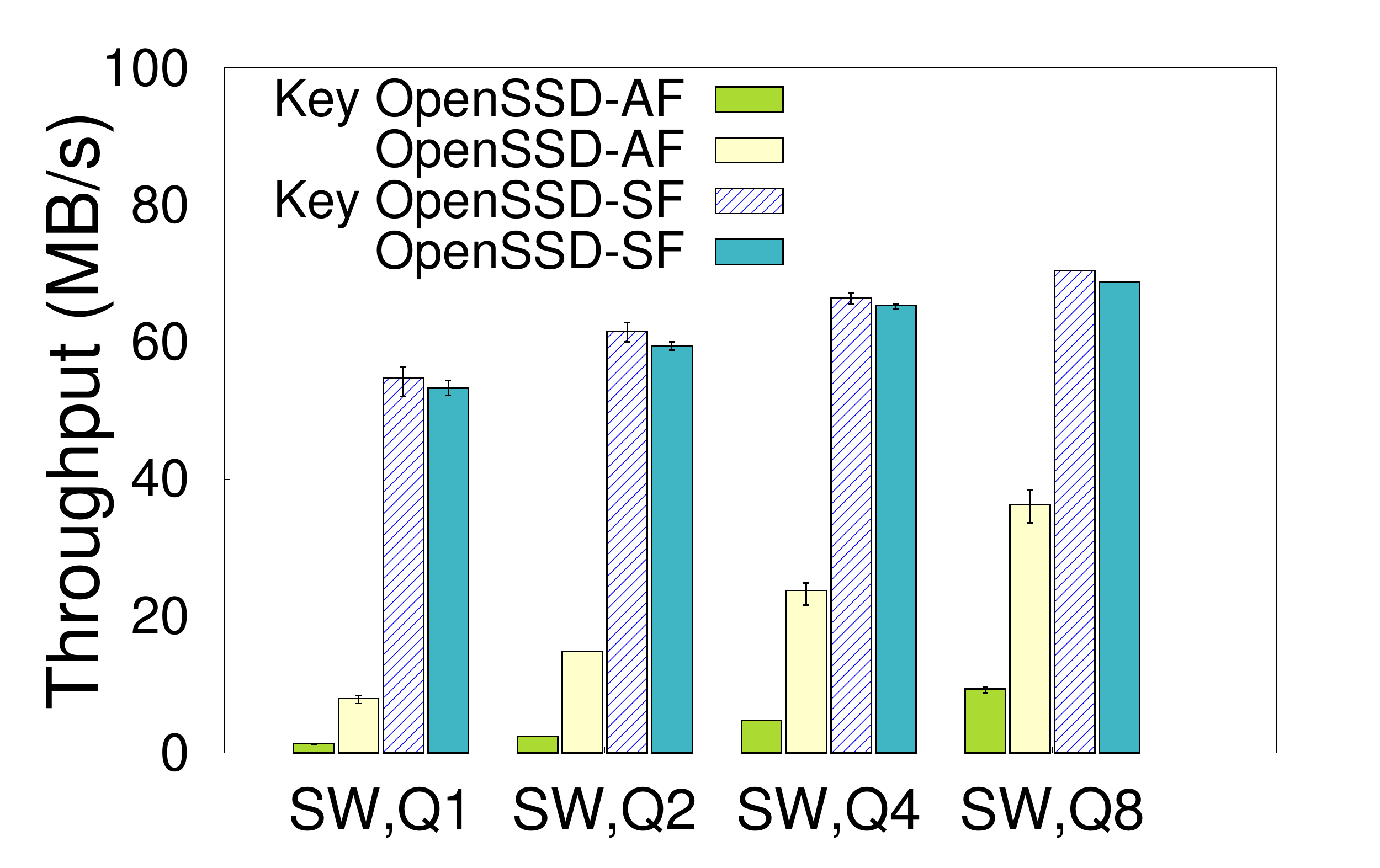} & 
			\includegraphics[width=0.25\textwidth]{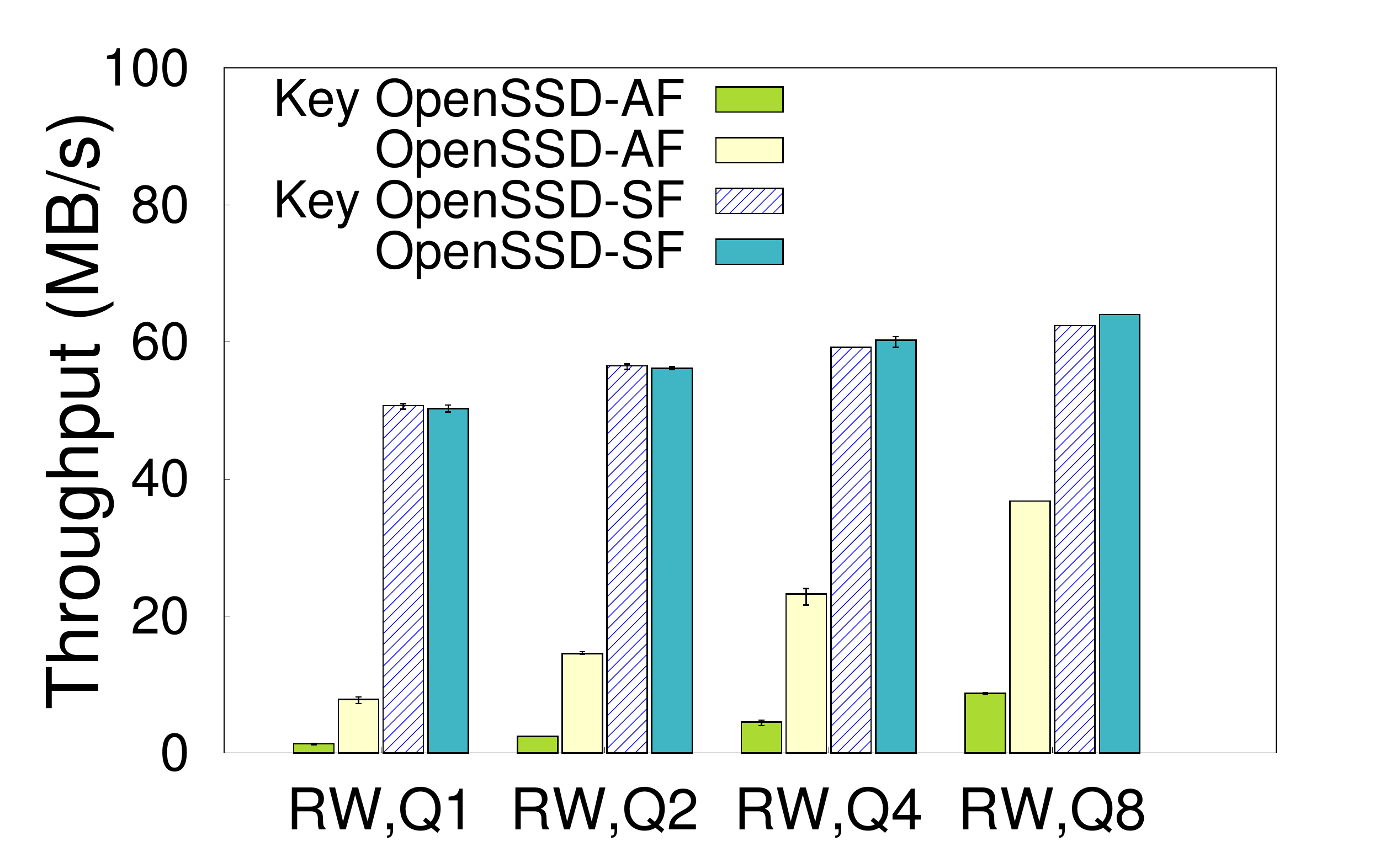} &
			\includegraphics[width=0.25\textwidth]{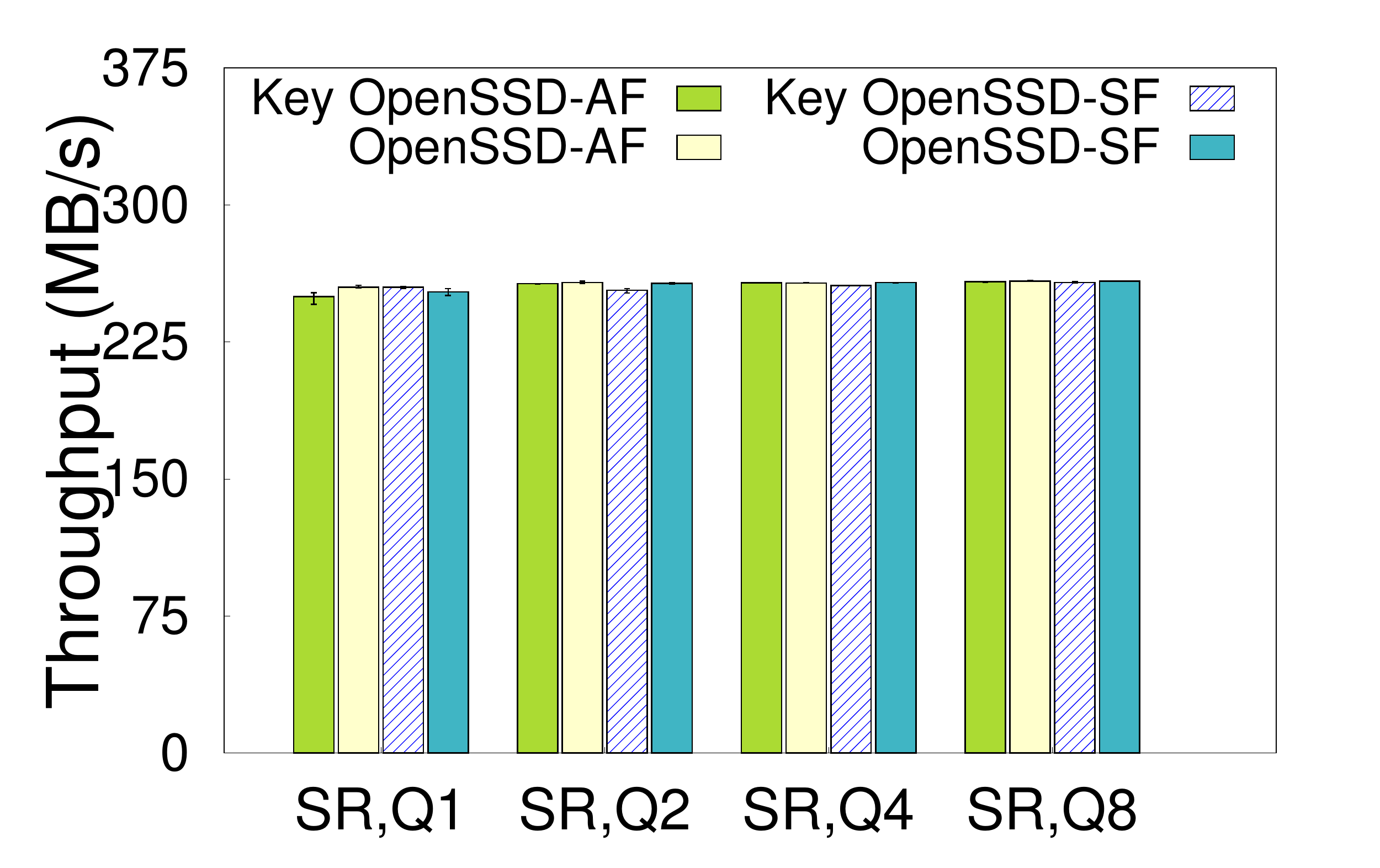} & 
			\includegraphics[width=0.25\textwidth]{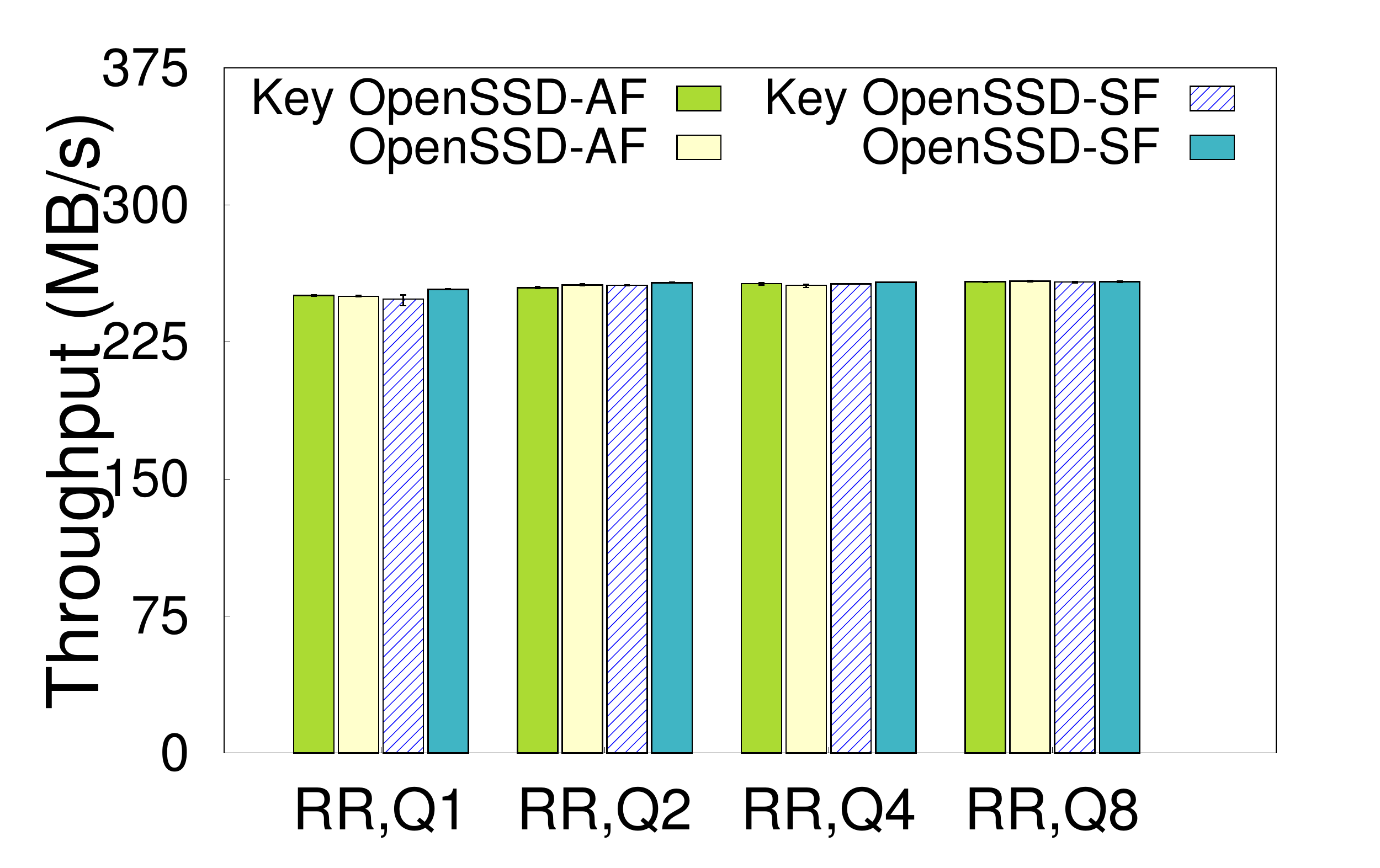} \\
						(a) Sequential Write & (b) Random Write &
 (c) Sequential Read & (d) Random Read\\
		\end{tabular}
		\vspace{-0.15in}
		\caption{Performance comparisons of {\statickeyftl} in Key OpenSSD and normal FTL in OpenSSD for 
		write and read workloads. 
SW, RW, SR, and RR denote Sequential Write, Random Write, Sequential Read and Random Read respectively. Queue depth is denoted Q$i$ where $i$ is the number of outstanding I/O requests. 
		AF and SF denote All Flush and Selective Flush respectively.}
		\label{fig:key-FTL-eval}
		\vspace{-0.35in}
	\end{center}
\end{figure*}

{\bf Test-bed:}
All experiments were performed on a single server with 16 GB of RAM and an Intel (R) Core (TM) i5-7500 CPU @ 3.40GHz. The operating system is Linux with kernel 4.10.16.
We examined two storage devices that are detailed in Table~\ref{tab:ssd_spec}. 
We selected the Micron 250GB MLC SSD as a baseline to measure Linux kernel implementation overhead for {\keyssd}. In order to include the implementation overhead of internal FTL and the SATA target driver on the SSD, 
we modified the firmware of the Jasmine OpenSSD development platform~\cite{jasmine}.

{\bf Workloads:}
We examined the key transmission and processing overhead on the SSD in terms of I/O bandwidth. To measure the overhead of disk-side implementation, we used an I/O Benchmark Suite, fair-lio~\cite{olcf_benchmark_suit} that uses the \texttt{libaio} asynchronous I/O library on Linux~\cite{libaio}, 
performing reads and writes on raw block devices. 
We have also made an in-house multi-process based I/O benchmark program to evaluate the overhead of Linux kernel implementations for key transfer and management 
in the kernel. 
To evaluate {\keyssd}, this benchmark program use \texttt{sys\_open\_key()} and \texttt{sys\_close\_key()} to pass keys to the kernel during I/Os.
We used two representative file sizes to have different file groups
because the data center workload consists of many small files and a few large files~\cite{DBLP:conf/fast/KimAVS15}. 
We used 4~KB {\em small files} 
and 
512~MB {\em big files}.
We also used SQlite~\cite{sqlite} and DBbench~\cite{dbbench} for the end-to-end performance analysis of {\keyssd} for realistic experiments. 
An actual ransomware sample is used to show that
{\keyssd} can prevent from both normal file I/O path attack and Direct I/O attack bypassing file system layers in the OS. 
In particular, for every experiment to measure the performance overhead of kernel and FTL implementations, 
we did a page cache flush to rule out the OS page cache effect .

\begin{table}[!t]
	\centering
	\scriptsize
	\begin{tabular}{|M{30mm}||M{12mm}|M{20mm}|M{4mm}|M{4mm}|M{6mm}|} \hline
		\textbf{SSD Spec} & \textbf{Jasmine} & \textbf{MX200 SSD} \\  \hline\hline
		{Company} 	& Indilinx	& Crucial  Micron \\ \hline
		{Type} 		& MLC	& MLC  \\ \hline
		{Interface} 	& SATA 	& SATA \\ \hline
		{Capacity} 	& 64~GB	& 250~GB \\ \hline
		{Read (MB/s)} 	& 270 & 555 \\ \hline
		{Write (MB/s)} 	& 90 	& 556 \\ \hline
	\end{tabular}
	\vspace{-0.1in}
	\caption{SSD Specification.
	Read and write bandwidth was measured using I/O Benchmark Suite~\cite{olcf_benchmark_suit}.
	}
	\label{tab:ssd_spec}
	\vspace{-0.2in}
\end{table}
 

{\bf Threat Model:}
According to our threat model in Section~\ref{sec:back}, a system call may be hijacked. If an application sends a plain key to OS, it could be revealed by a hijacked system call. Since the focus of this paper is on demonstrating the effectiveness of integration of an access-control mechanism with a disk drive, we used a plain key in our current implementation. 
But, a signature that is encryption of a hash value of the address and data of each request can be used instead of the plain
key. 

\vspace{-0.15in}
\subsection{Overhead Analysis of {\keyftl}}
\vspace{-0.05in}

In this section, we show the overhead analysis of {\statickeyftl} in {\keyssd} versus normal page-based FTL in SSD. 
Firstly,
we compared the performance of I/O hardware and software queues using the fair-lio I/O benchmark suite. 
We observed that there is negligible performance difference between hardware and our software queue implementations, but owing to space constraints here, we do not show results. 

Secondly, we evaluated the disk-level performance overhead of {\statickeyftl} for {\keyssd}. We compared the write and read I/O performance of {\keyssd} with {\keyftl}, called {\em Key OpenSSD} and normal SSD with normal FTL (with no key), called {\em OpenSSD} for a variety of I/O workload patterns (sequential and random) by varying I/O queue depth. 
Figure~\ref{fig:key-FTL-eval}(a)(b) show the results to compare {\keyssd} with {\statickeyftl} and normal SSD for write only workloads. 
We evaluated the performance comparison by changing the FTL flushing methods ({\em all flush (AF)} or {\em selective flush (SF)}). 
Note that all flush means all entries of the mapping table are all synchronized no matter what entries are changed while selective flush means only changed entries from the table are synchronized. 
In case of all flush, it is observed that the throughput of {Key OpenSSD} 
is less than 20-30\% of that of {OpenSSD}.
It is because the size of the {\statickeyftl} in Key OpenSSD is 
nine times larger than that of the normal FTL. A page size in OS is 4~KB whereas a page size in the Jasmine OpenSSD is 32~KB. Thus, an entry of FTL has eight key values for each 4~KB page. 
However, in case of selective flush, we observe that there is little difference between Key OpenSSD and OpenSSD due to small flush overhead. 
Figure~\ref{fig:key-FTL-eval}(c)(d) show the results for read only workloads. 
Read workloads rarely flush, so there is little difference in performance between {\keyssd} and normal SSDs.

\begin{figure}[!t]
	\begin{center}
		\begin{tabular}{@{}c@{}c@{}c@{}c@{}}
			\includegraphics[width=0.25\textwidth]{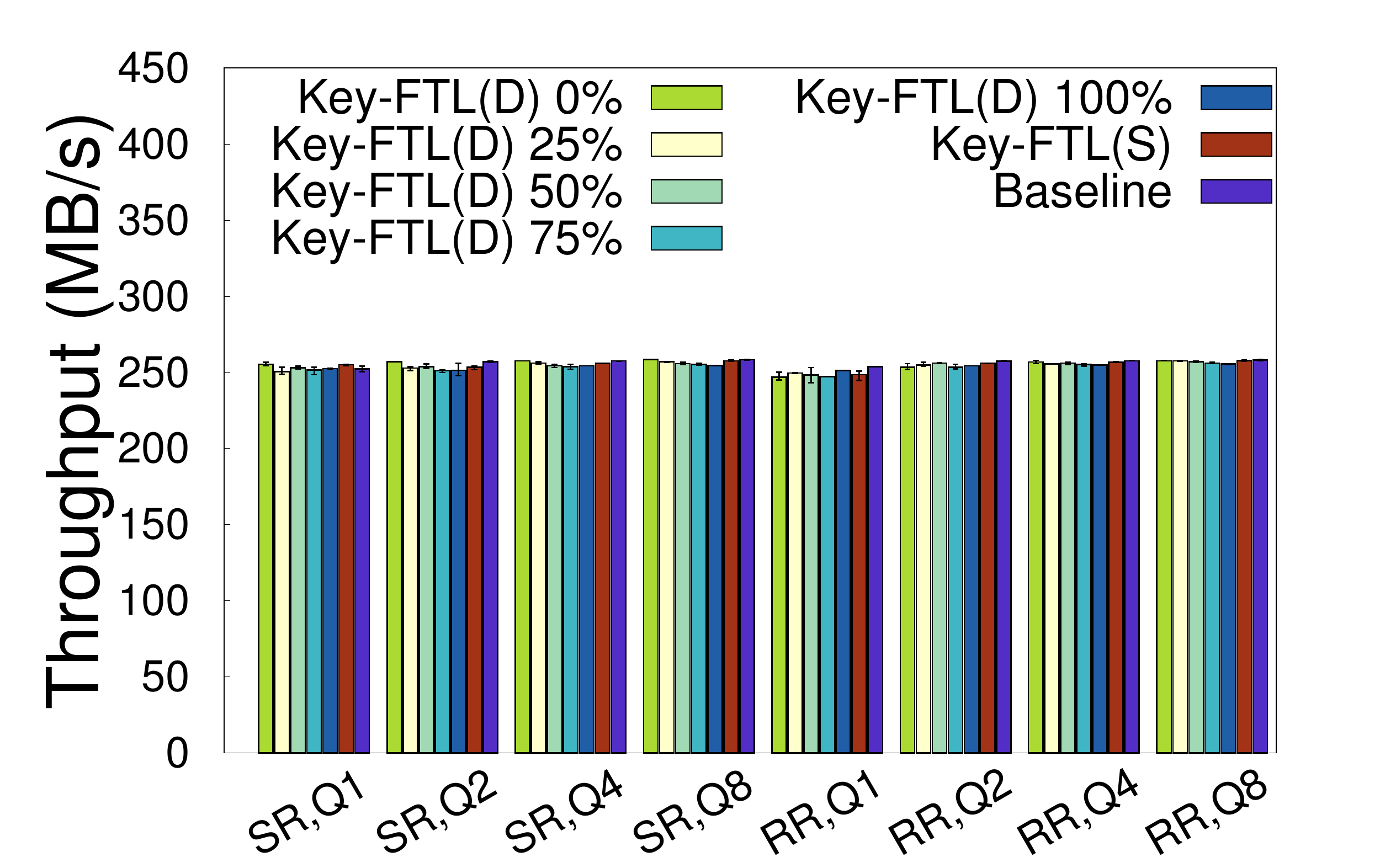} & 
			\includegraphics[width=0.25\textwidth]{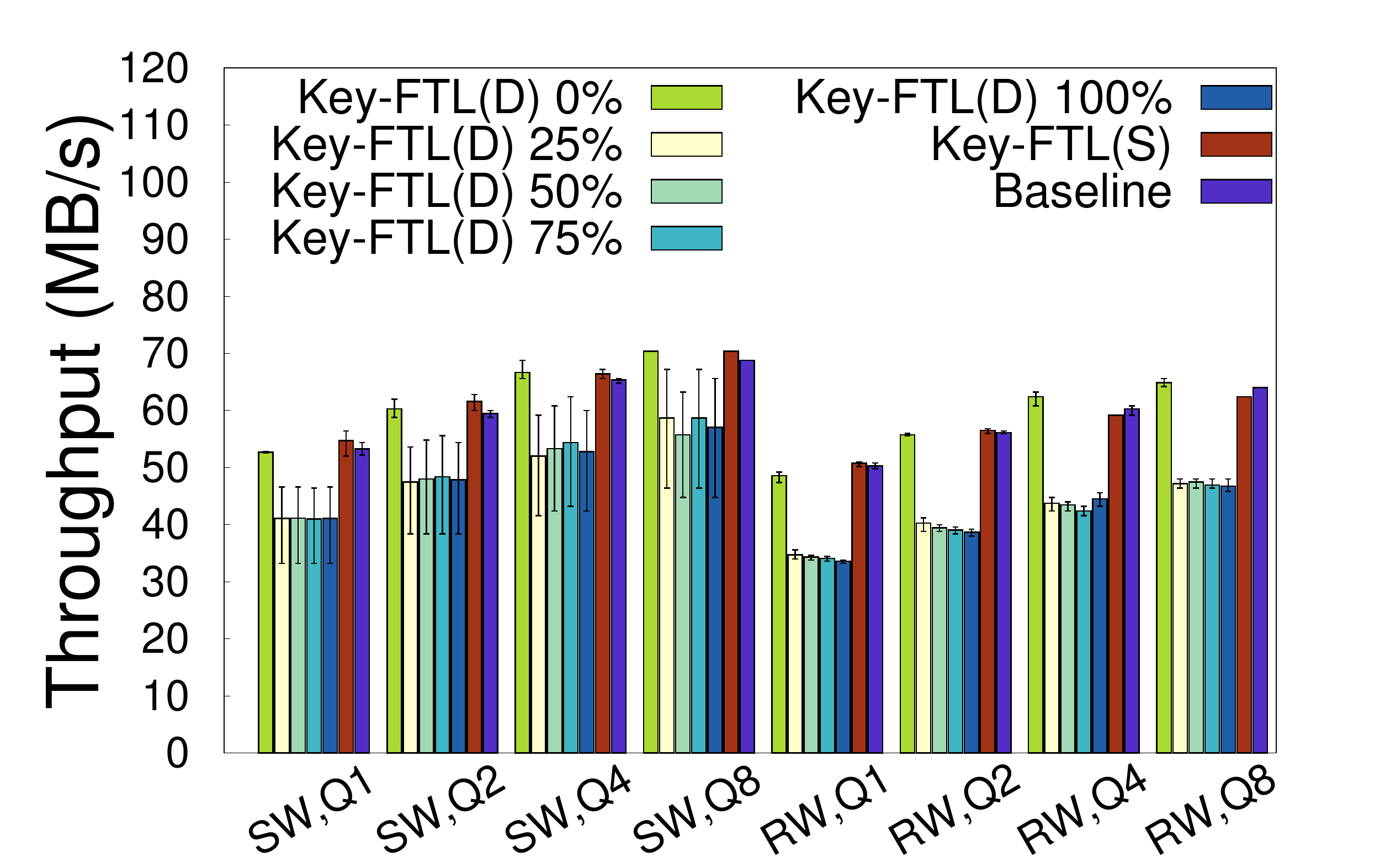}\\										(a) Read & (b) Write\\
			\vspace{-0.3in}
		\end{tabular}
		\caption{Performance analysis of {\dynamickeyftl} by varying the percentage of blocks locked by keys. }
		\label{fig:static_and_dynamic}
		\vspace{-0.3in}
	\end{center}
\end{figure}

{\bf Performance analysis of {\dynamickeyftl} and {\statickeyftl}:}
In {\dynamickeyftl}, a key is dynamically allocated. It is advantageous in terms of space compared to {\statickeyftl}, but the performance of {\dynamickeyftl} will depend on the portion of blocks locked by keys. 
For performance evaluation, we partitioned 4GB and measured  direct I/O performance by varying the percentage of blocks locked among all blocks to 
0\% (0GB), 25\% (1GB), 50\% (2GB), 75\% (3GB), and 100\% (4GB).
We ran the fair-lio I/O benchmark suite for a variety of I/O patterns such as sequential read and write and random read and write by increasing queue depth. 
0\% means there is no protected blocks by keys, which is a similar case to a baseline FTL without keys. On the other hand, 100\% means all data blocks are protected by keys. 

Figure~\ref{fig:static_and_dynamic}(a) shows the results for reads. We observe read bandwidth is around 250 MB/s regardless of the percentage of locked blocks. It means performance is almost equivalent to the baseline without keys and {\statickeyftl}. 
{\dynamickeyftl} has a search overhead for granting access to the LPNs corresponding to the key, however, with our results, the search overhead can be said almost negligible. 
Figure~\ref{fig:static_and_dynamic}(b) shows the results for writes. 
The baseline without keys and {\statickeyftl} show very similar performance while 
{\dynamickeyftl} shows decreased throughputs.  
This is because the overhead of inserting key nodes is larger than the search overhead. In our implementation for {\dynamickeyftl},
it is necessary to register the key in {\keyftl} and lock it for all LPNs of the request sent from the host. 
However, these results are for worse case when all LPNs are for first writes. 
If they are update-writes on blocks already locked, it only involves the search overhead of LPNs for granting access with the key, so performance will get better.

\begin{figure*}[!t]
	\begin{center}
		\begin{tabular}{@{}c@{}c@{}c@{}c@{}}
			\includegraphics[width=0.25\textwidth]{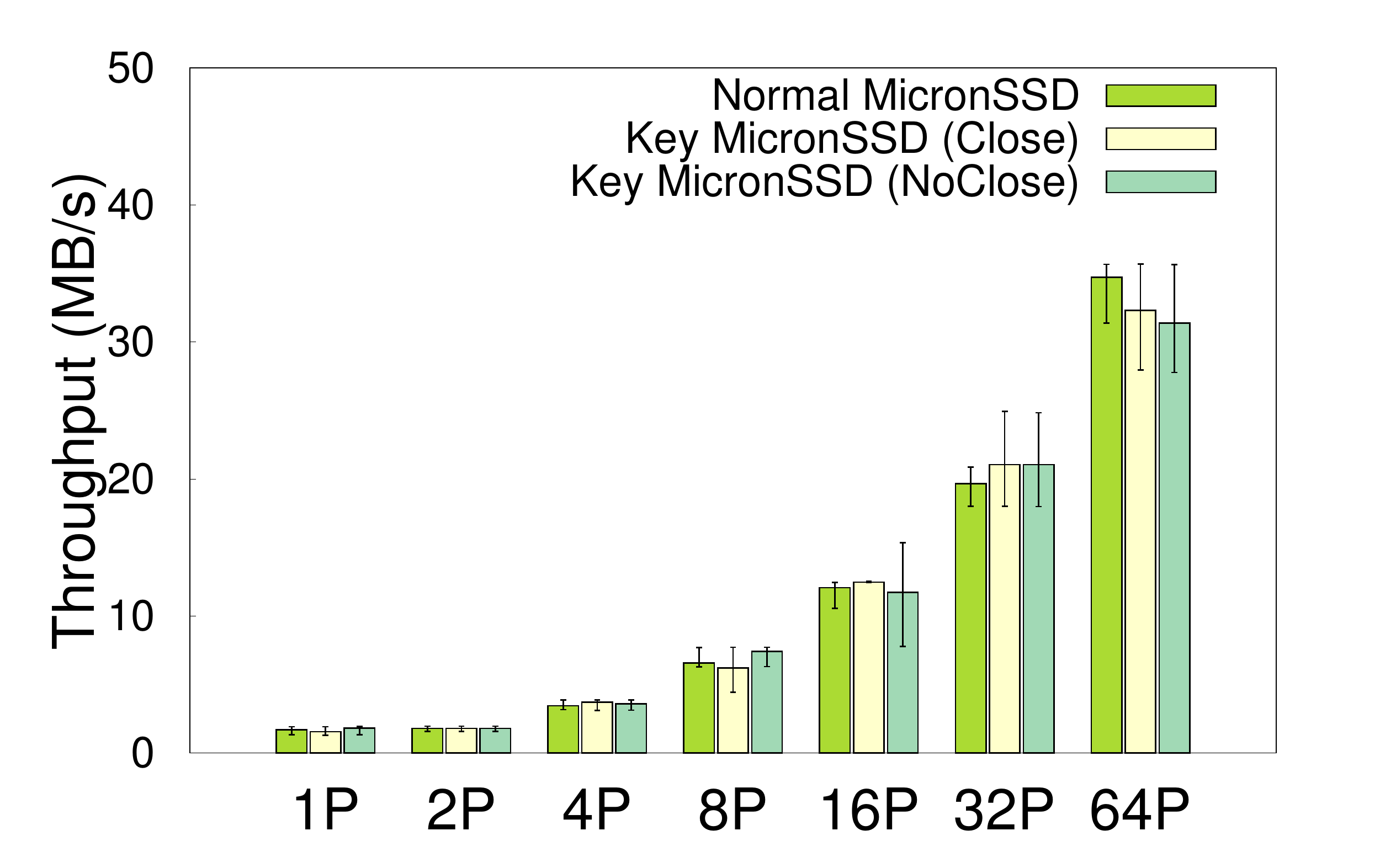} & 
			\includegraphics[width=0.25\textwidth]{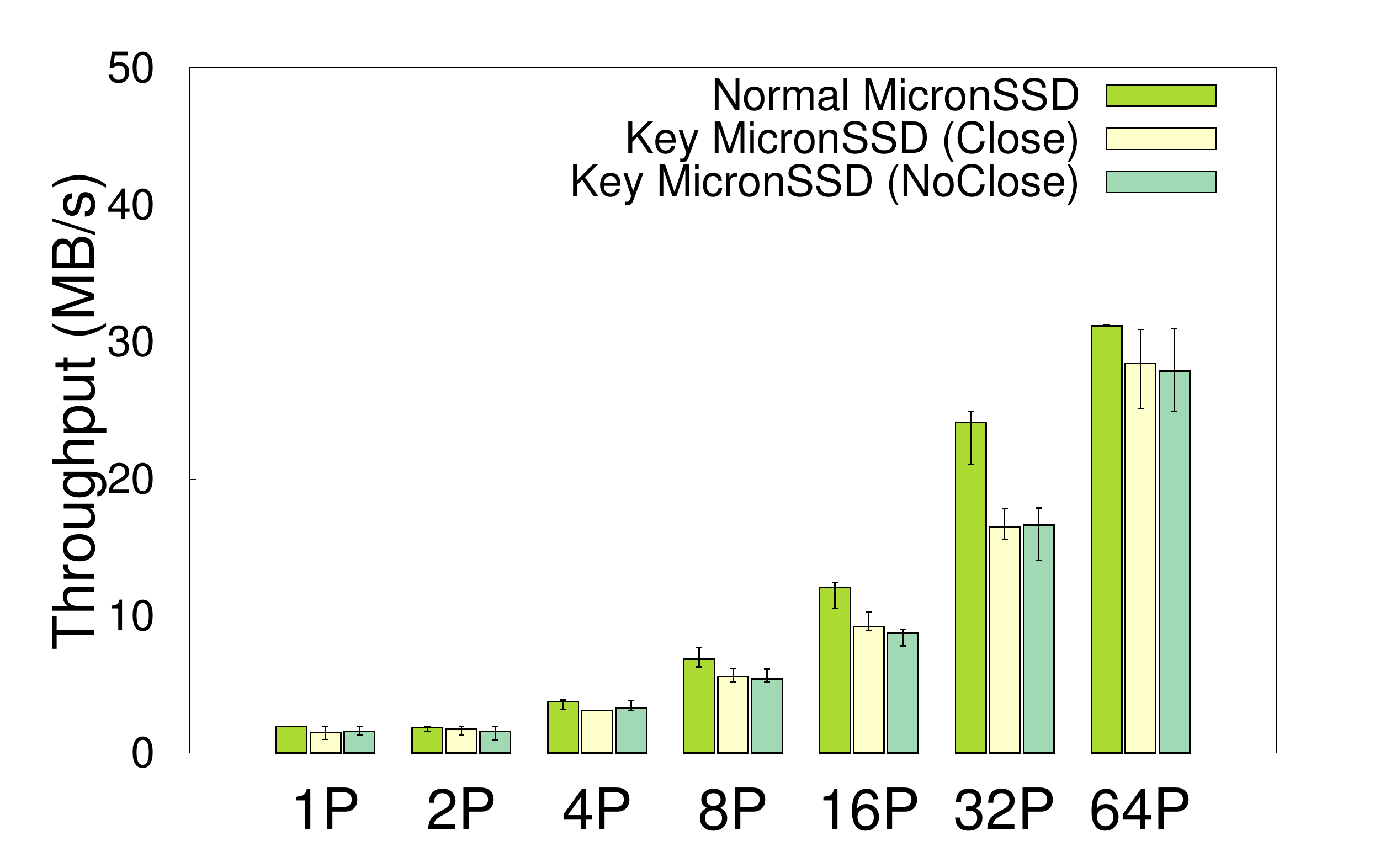} &
			\includegraphics[width=0.25\textwidth]{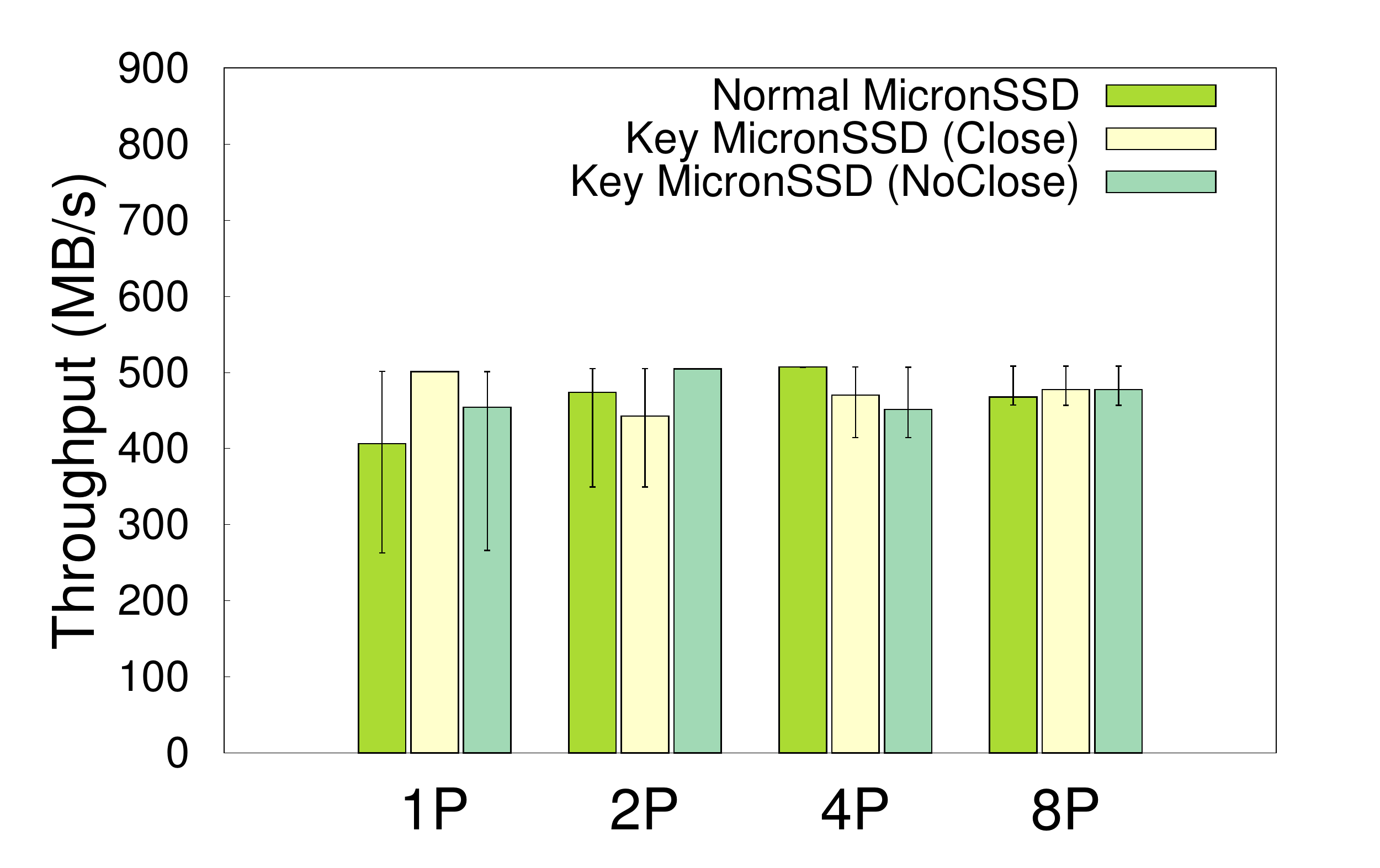} & 
			\includegraphics[width=0.25\textwidth]{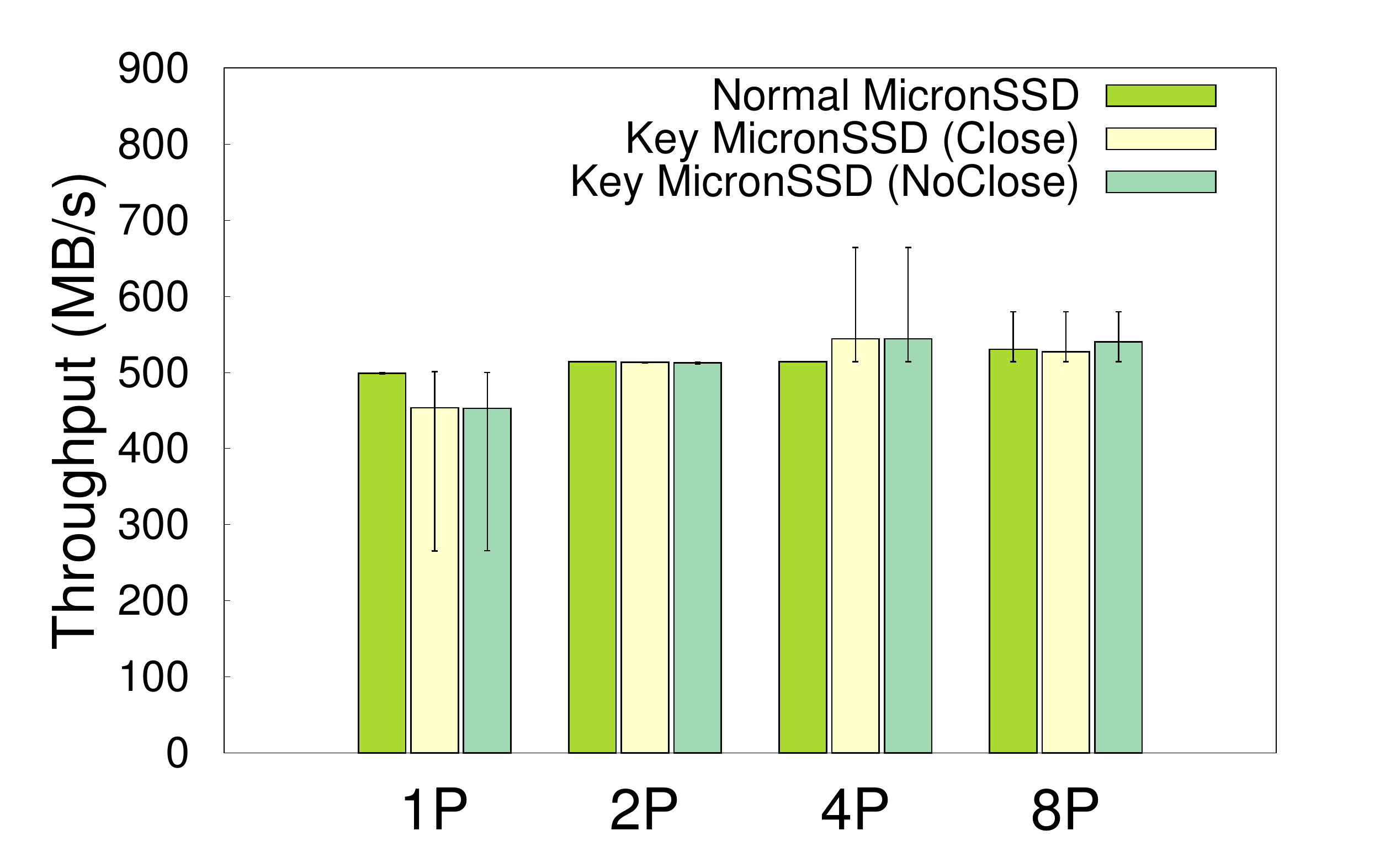} \\
			(a) Write (Small Files) & (b)  Read (Small Files) & (c) Write (Big Files) & (d)  Read (Big Files) \\
					\vspace{-0.25in}
		\end{tabular}
		\caption{Analysis of the kernel implementation overhead for {\keyssd} with Micron MLC SSD.
		Small Files and Big Files in the parenthesis denote workload type according to file size.
		The error bar depicts min. and max. deviation from the average of 3 iterations.
		}
		\label{fig:kernel_overhead_ssd}
		\vspace{-0.25in}
	\end{center}
\end{figure*}

\begin{figure*}[!t]
	\begin{center}
		\begin{tabular}{@{}c@{}c@{}c@{}c@{}}
			\includegraphics[width=0.25\textwidth]{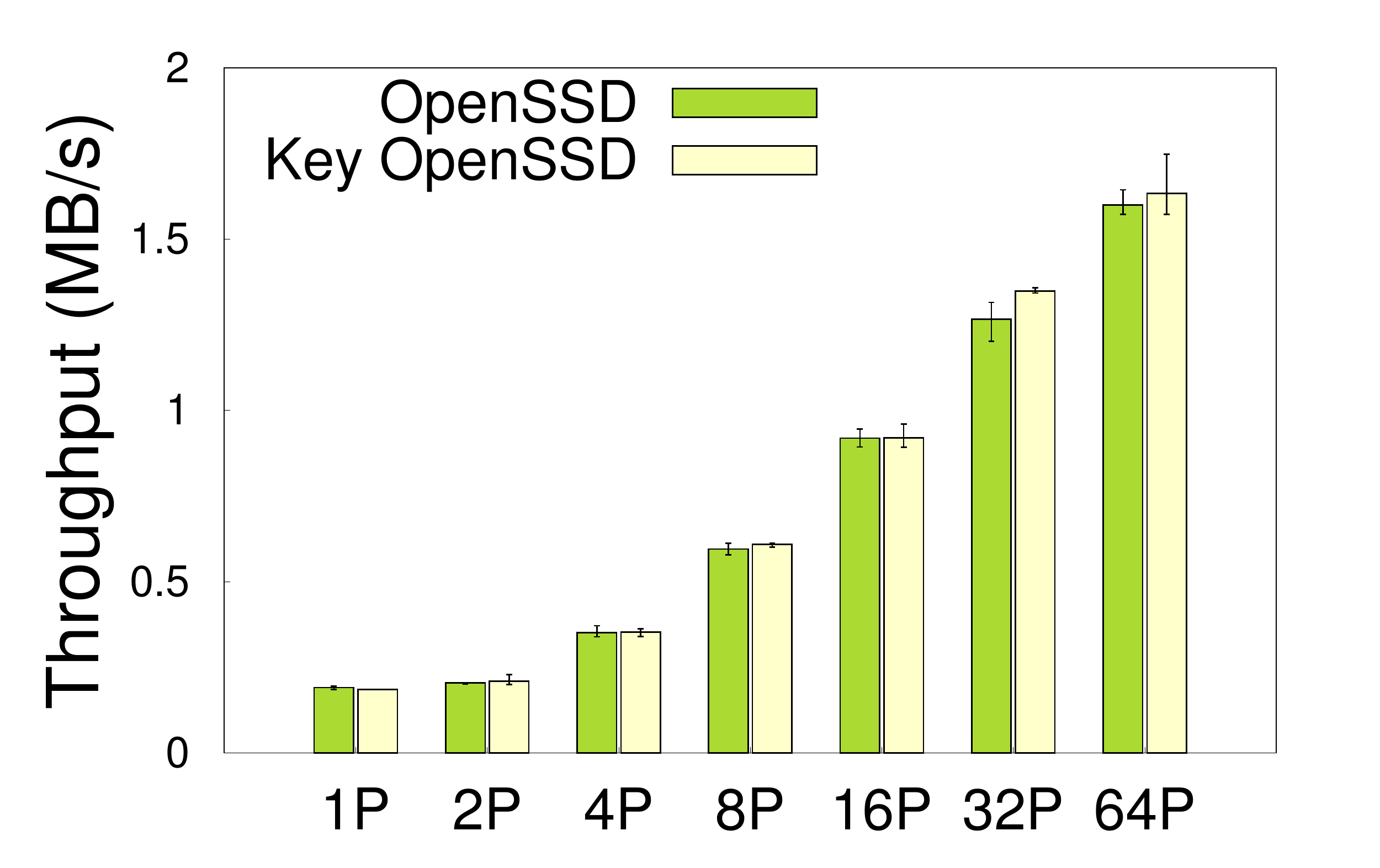} & 
			\includegraphics[width=0.25\textwidth]{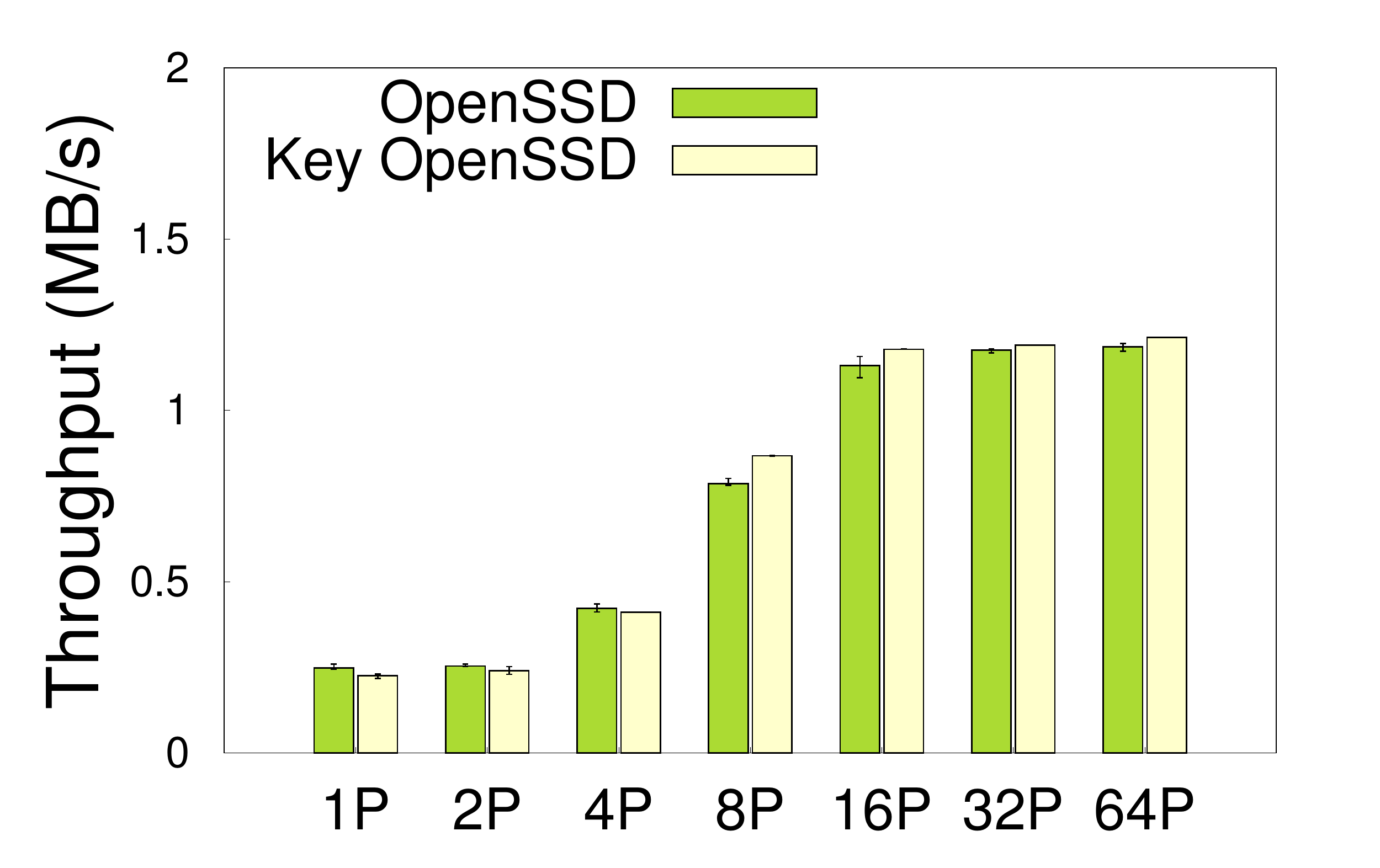} &
			\includegraphics[width=0.25\textwidth]{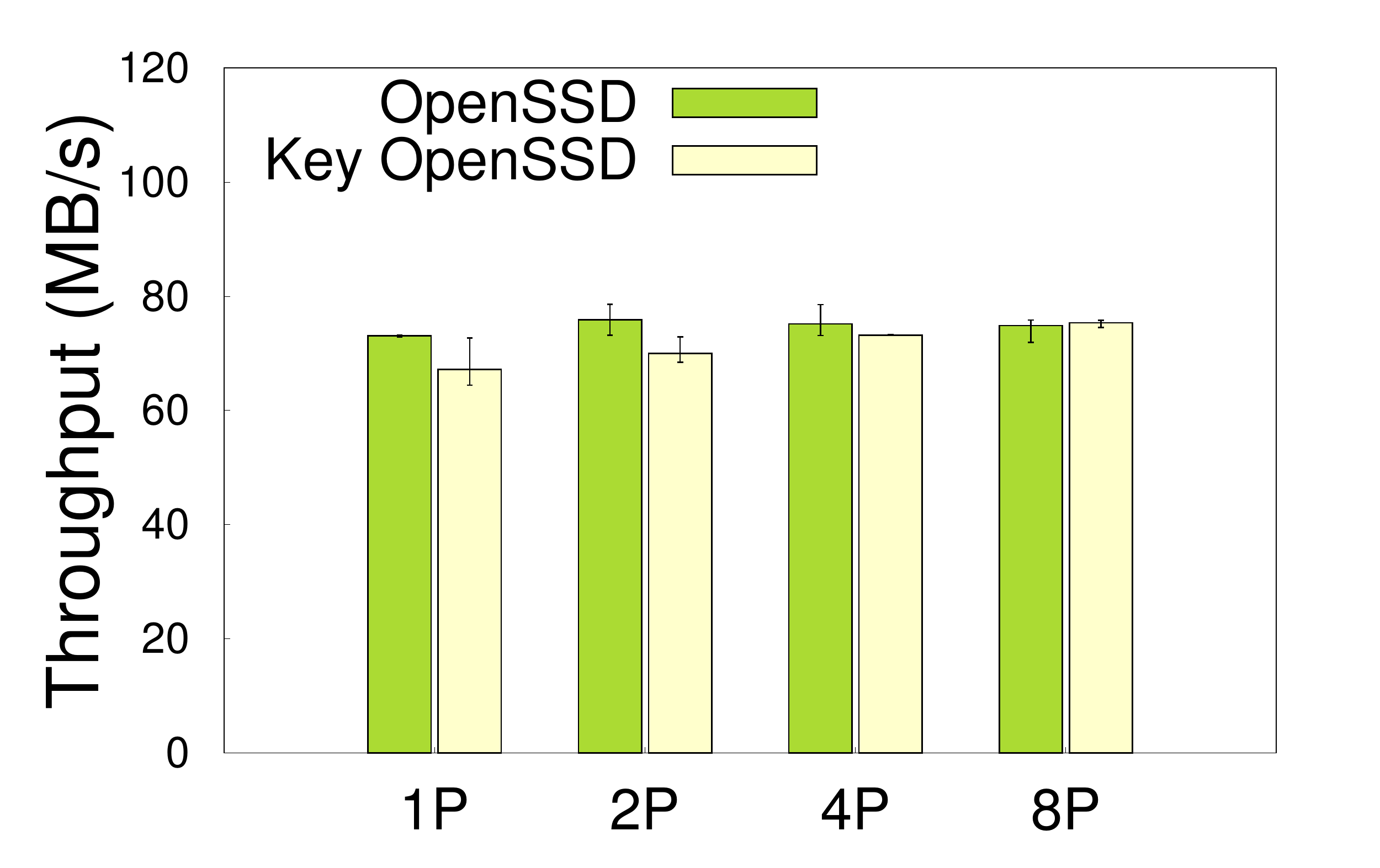} & 
			\includegraphics[width=0.25\textwidth]{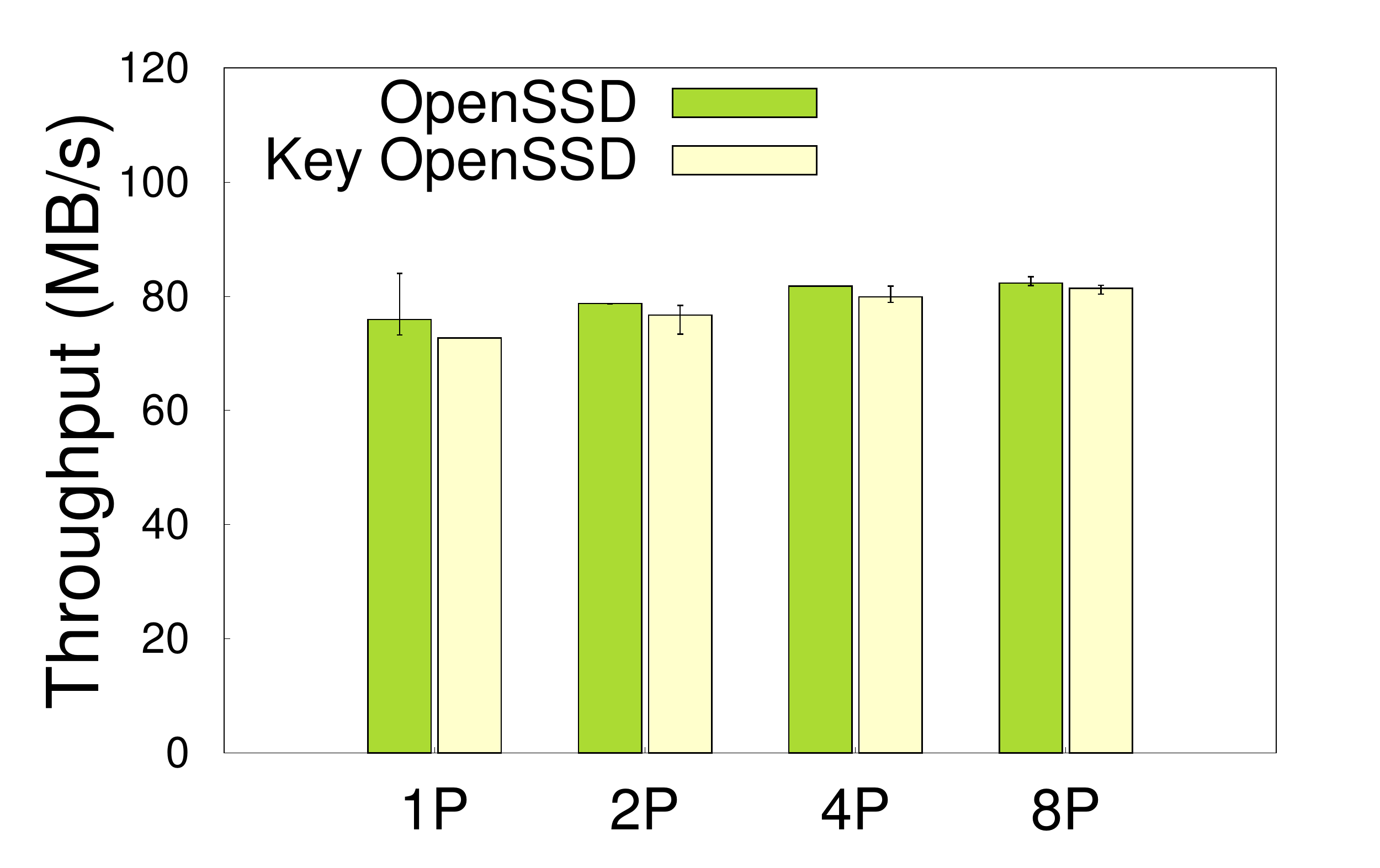} \\
			(a) Write (Small Files) & (b)  Read (Small Files) &
			(c) Write (Big Files) & (d) Read (Big Files)\\
					\vspace{-0.25in}
		\end{tabular}
		\caption{
		Analysis of the end-to-end performance experiment including the kernel and firmware implementation overhead using the Jasmine OpenSSD. 
		 The experimental environment is the same as that for Figure~\ref{fig:kernel_overhead_ssd}.
		}
		\label{fig:kernel_overhead_jasmine}
		\vspace{-0.3in}
	\end{center}
\end{figure*}

\vspace{-0.15in}
\subsection{Kernel Implementation Overhead}
\vspace{-0.05in}

In this experiment, we analyze the kernel implementation overhead for key transmission and management for {\keyssd}. 
The experiments were performed by increasing the number of processes, each generating the same workload.
Figure~\ref{fig:kernel_overhead_ssd}
compares the performance results of {\keyssd} (Key MicronSSD) with normal SSD (Micron SSD). 
This experiment was performed using a Micron MLC SSD to analyze only the kernel implementation overhead, excluding the {\keyftl} implementation overhead. 
The SATA driver sends the key in the Register FIS, but the SSD ignores the key.
In particular, in order to analyze the performance cost of key search in {\keyinode}, we performed two experiments -- (i) to delete key from {\keyinode} when closing (Close) the file and (ii) that does not delete key when closing the files (NoClose).

From Figure~\ref{fig:kernel_overhead_ssd}, 
we observe the results for small files show lower throughputs than big files workload. 
Small workload performs I/O operations on a large number of small files and involves large delays caused by opening and closing files each time for many files, as opposed to doing continuous I/Os on large files with large workloads.
From Figure~\ref{fig:kernel_overhead_ssd}(a)(b), we observe increased throughput as the number of processes increases in all three cases (MicronSSD, Key MicronSSD (Close), Key MicronSSD (NoClose)). 
Comparing the results for MicronSSD and Key MicronSSD (Close), 
{\keyssd} (Key MicronSSD) seems to have slightly lower throughput than  MicronSSD when I/O loads are high (referring to 32P or 64P). 
Since the OS kernel runs the read-verify method on the {\keyssd}, it requires additional disk block access for key authentication on the first page when opening the file.
When we compare the results for Key MicronSSD (Close) and Key MicronSSD (NoClose), 
we again see the performance difference between the two is almost negligible, unlike the expectation that the key retrieval time will take longer for all key searches because the keys are kept in the {\keyinode} table without being deleted.
 
For example, in Key MicronSSD (NoClose), I/O is performed for 64,000 files in 64 processes. If the key is not removed from the {\keyinode} at the time of file close, up to 64,000 entries may accumulate in the {\keyinode}, which may degrade table search performance. However, in our experiment, the search time overhead is too small to degrade the overall I/O performance. 
Figure~\ref{fig:kernel_overhead_ssd}(c)(d) show the results for big files workloads. 
Unlike the results for small workloads, we see there is very little performance difference between three. This is because the additional disk access overhead of the read-verify method is not noticeably large.

\vspace{-0.15in}
\subsection{End-to-End Performance Analysis}
\vspace{-0.05in}

In this experiment, we perform an end-to-end experiment to analyze the overhead including the kernel and 
{\statickeyftl} implementation overhead.
For the Key OpenSSD, we deleted keys from 
{\keyinode} when closing.
Figure~\ref{fig:kernel_overhead_jasmine} shows the performance comparison between OpenSSD and Key OpenSSD. 
Overall bandwidths were observed lower than those from Micron experiments because overall read/write bandwidths of OpenSSD are lower than MicronSSD as in Table~\ref{tab:ssd_spec}.
Comparing the performance of OpenSSD and Key OpenSSD, we see little difference in performance between them, which is slightly different from 
Figure~\ref{fig:kernel_overhead_ssd}. 
In particular, in Figure~\ref{fig:kernel_overhead_ssd}, 
the kernel overhead caused by the read-verify method is noticeable when I/O loads are very high, but it is not shown here. This is because the performance of the OpenSSD is so low that the overhead of the read-verify method is hidden.


For another realistic end-to-end experiment, we have modified the SQLite~\cite{sqlite} source code and protected DB files with a key. 
SQLite could pass a user-defined key when accessing DB files using \texttt{sys\_open\_key()} and \texttt{sys\_close\_key()} system call. 
Table~\ref{tab:sqlite_test} presents the results of comparing the performance of OpenSSD and Key OpenSSD. 
Performance was measured using the Database Benchmark Tool~\cite{dbbench} in SQLite~\cite{sqlite} with two representative DB workloads. 
Insert workload is write heavy and intersection workload is read heavy. 
For the insert workload, we observe OpenSSD shows 16.48 TPS (Transactions Per Second) while Key OpenSSD shows 16.42 TPS, which is little difference between them. For the intersection workloads, we have similar observation. 

\begin{table}[!t]
	\centering
	\footnotesize
	\begin{tabular}{|M{20mm}||M{14mm}|M{20mm}|M{4mm}|M{4mm}|M{6mm}|} \hline
		\textbf{Workload} & {\bf Key OpenSSD} & {\bf OpenSSD}\\  \hline\hline
		{Insert} 	& 16.42	& 16.48 \\ \hline
		{Intersection} 		& 880.80	& 892.23  \\ \hline
	\end{tabular}
	\vspace{-0.05in}
	\caption{Performance comparison of OpenSSD and Key OpenSSD for database application. 
	}
	\label{tab:sqlite_test}
	\vspace{-0.2in}
\end{table}

\begin{figure*}[!t]
	\begin{center}
		\begin{tabular}{@{}c@{}c@{}c@{}c@{}}
			\includegraphics[width=0.96\textwidth]{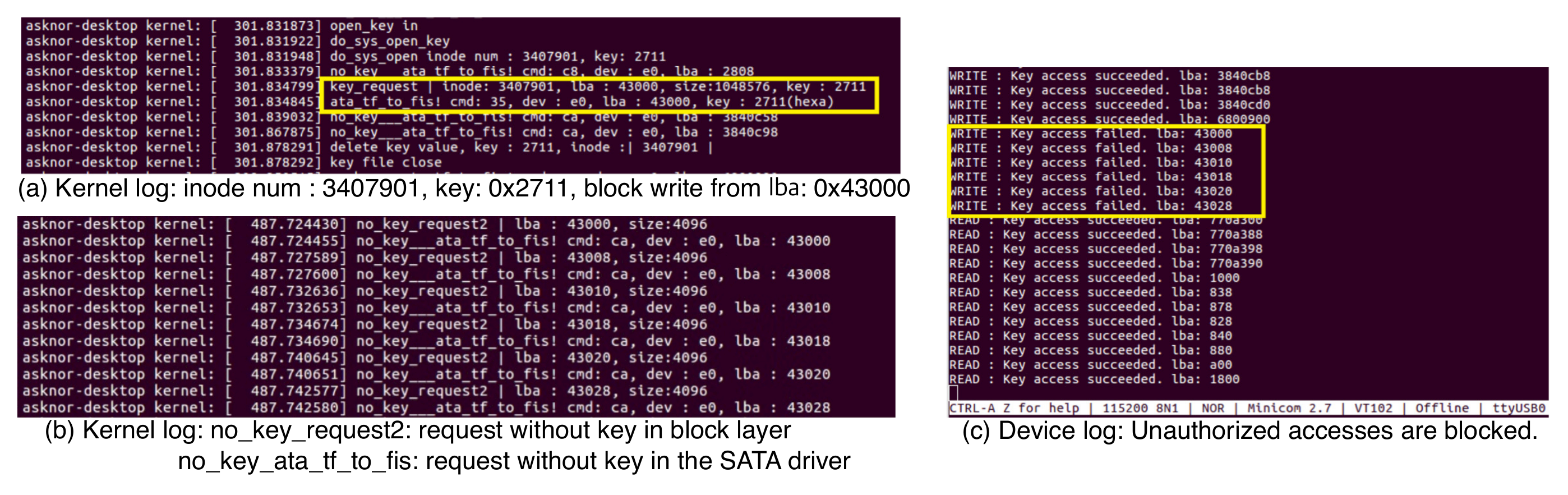}  \\
			\vspace{-0.35in}
		\end{tabular}
		\caption{
		Verification of protection by {\keyssd} for direct I/O ransomware attacks.
		}
		\label{fig:direct_io}
		\vspace{-0.35in}
	\end{center}
\end{figure*}

\vspace{-0.15in}
\subsection{Protection against Real Ransomware}
\vspace{-0.05in}

In this experiment, we show {\keyssd} can protect against attacks performed through the normal file I/O path of the OS and by bypassing the OS directly from the user application.

{\bf Normal file I/O path attack:}
To evaluate whether {\keyssd} can prevent actual ransomware attacks, we run an ransomware sample~\cite{ransomware} on the SQLite DB file (test.db) generated in the previous experiment. 
The ransomware sample works as follows.
The ransomware first reads test.db, the target file, and encrypts it with AES 256 encryption. Then, it creates a new file, infected.db and write the encrypted data on it, and then deletes the original db file, test.db. 
Ransomware opens test.db without a key. 
When attempting to read data (test.db), ransomware does not have a key, so the block request can not find the correct key when referring to the {\keyinode} in the kernel block layer, and eventually the access is blocked at the disk end.

{\bf Direct I/O attack:}
We formatted OpenSSD as EXT4 and created a 1M file with a key. We also used a fair-lio benchmark to directly access the device and perform I/O on the data blocks on the disk. Then, we write to the 1MB file data locked by the key on the disk in 4 KB units without the key, and write 24 KB sequentially.
Figure~\ref{fig:direct_io} shows the host's kernel log and OpenSSD's device log.
Currently a 1M file is written beginning with LBA: 0x43000.
Figure~\ref{fig:direct_io}(a) shows the kernel and OpenSSD device logs on the host for a 24 KB write request without a key.
Figure~\ref{fig:direct_io}(b)(c) show the kernel and device's logs when attempting to write on the locked blocks with the proper key. 
Figure~\ref{fig:direct_io}(b) shows kernel logs, 
kernel block layer and device driver send the 24 KB write requests without keys.
And Figure~\ref{fig:direct_io}(b), Figure~\ref{fig:direct_io}(c) shows {\keyssd} 
blocks LBA's requests on the blocks from 43000 to 43028.
Therefore, this experiment confirmed that {\keyssd} protects direct I/O access of unauthorized applications. 

{\bf Attacks using page cache:}
There could be a ransomware attack that reads the file data in the page cache and deletes the original file after creating a new file. 
As mentioned in Section~\ref{sec:issue}, {\keyssd} can prevent these attacks through the read-verity method. We have experimentally proved that this attack can be successfully defended. However, owing to space constraints here, we do not show results.

\input{eval_security_anal}

%% file: eval_security_anal.tex
\vspace{-0.15in}
\subsection{Security Analysis}
\label{sec:sec_anal}
\vspace{-0.05in}

In this section, we discuss the steps that ransomware typically performs and the effectiveness of Key-SSD, explaining how Key-SSD can defend against ransomware attacks at each step. The steps are as follows: Infection $\rightarrow$ Persistence $\rightarrow$ Removing backup copies $\rightarrow$  Encryption $\rightarrow$  Notice, illustrated in Table~\ref{tab:security_anal}.


{\bf Infection:}
Ransomware is a type of malware that exploits vulnerabilities to damage victim's computers. 
For example, CryptoLocker exploits vulnerabilities in Internet Explorer and Adobe Flash to control the computer~\cite{7886569}. 
The approach proposed in this study is that the application provides the key.
If the application is compromised, the files of the compromised application can be infected by the ransomware.
However, its impact is confined only 
to 
the compromised application.
If an application is compromised, its key 
is very likely to
be revealed.
In this case, files belonging to the compromised application 
can be encrypted by the ransomware. 
However, since the revealed key is only for the compromised application, ransomware cannot use the key to access other files that belong to other applications.
For example, even if CryptoLocker compromises Internet Explorer, \texttt{docx} files and \texttt{pptx} files that belong to MS Word and PowerPoint cannot be accessed by CryptoLocker.

{\bf Persistence:}
Ransomware can survive after rebooting.
It may also include a self-propagating mechanism that infects other computers in the network.
For example, WannaCry propagates itself by exploiting vulnerability of Window's Server Message Block (SMB) protocol~\cite{chen_robert}.
This persistence mechanism often 
modifies system files. 
Therefore, if system files 
are protected by {\keyssd},
it will be much harder for ransomware to modify system files.
For example, 
{\keyssd} can enforce system files to be updated only by the legitimate updater.

{\bf Removing backup copies:}
Ransomware destroys backup copies.
This step is unique to ransomware compared to other types of malware.
If backup copies are available, the victim can recover files without paying for ransom.
Thus, ransomware finds and destroys 
all backup copies.
Again, if backup copies are under the protection of 
{\keyssd}, ransomware will not be able to delete them. 

\begin{table}[!t]
	\scriptsize
	        \begin{tabular}{|M{28mm}@{}||M{11mm}@{}|M{13mm}@{}|M{4mm}|M{7mm}|} \hline
		\textbf{Step} & \textbf{Infection} & \textbf{Persistence} & \textbf{RBC}  & \textbf{Encrypt} \\  \hline\hline
		{Malware Migration~\cite{7778160,6620049}} & X &  &  &   \\ \hline
		{Renaming~\cite{7924925}} &  & X  &  &   \\ \hline
		{App. Behavior Monitor~\cite{7536529,7600214,7336353}} &  &  &  & X  \\ \hline
		{Net.  Behavior Monitor~\cite{7764294,7387902,7981522}} &  &  &   & X  \\ \hline
		{Crypto. Library~\cite{Kolodenker:2017:PDA:3052973.3053035,Lee2017b,Maiorca:2017:RAP:3019612.3019793}} &  &  &  & X  \\ \hline
		{File Backup~\cite{cldsafe_an_efficient_file_backup,Continella:2016:SSR:2991079.2991110}} &  &  &  & X  \\ \hline
		{ \cellcolor{gray!20}{\keyssd}} &   \cellcolor{gray!20}&  \cellcolor{gray!20}X &  \cellcolor{gray!20}X &  \cellcolor{gray!20}X  \\ \hline
	\end{tabular}
	\vspace{-0.1in}
	\caption{Step-by-step analysis from Infection to Encryption  of ransomware attacks and comparison with existing defense techniques and {\keyssd}.
	RBC denotes removing back copies.
	}
	\vspace{-0.2in}
	\label{tab:security_anal}
\end{table}

{\bf Encryption:}
Ransomware finds the data file and encrypts it
with an encryption key.
Before encrypting data files, ransomware (e.g. CryptoLock) 
can 
exchange encryption keys with a remote command and control (C\&C) server.
If data files are under the protection of 
{\keyssd},  
ransomware cannot read and write (overwrite)
the file 
unless 
the application of the file is compromised. 
Most (if not all) of the existing ransomware mitigation techniques work 
at
this step.
They detect ransomware by monitoring specific behaviors pertaining to key exchanges and encryption.
{\keyssd} prevents ransomware not only at this step, but also previous steps as explained above.
It should be noted that several prior works for 
malware mitigation techniques\cite{7778160,6620049}
investigate the problems of detecting and preventing ransomware 
at the first two steps because ransomware is also a kind of malware.

{\bf Notice:}
Eventually, ransomware informs the victim of ransom payment. Once it is received, ransomware will restore the hostage file and erase all forensic evidence.

%% file: related.tex
\section{Related Work}
\label{sec:related}
\vspace{-0.1in}

Since ransomware is a kind of malware, existing malware mitigation techniques can be used to detect and prevent ransomware~\cite{7778160}.
Ransomware exhibits specific behaviors such as 
searching target files, 
accessing 
cryptographic libraries
frequently, and exchanging encryption keys with a remote command-and-control 
servers.
There are studies 
to prevent ransomware by detecting these known 
behaviors~\cite{7536529,7600214,7336353, 7764294,7387902,7981522, Kolodenker:2017:PDA:3052973.3053035,Lee2017b}.
While these techniques increase the difficulty of successful infections, evolved ransomware can circumvent these techniques~\cite{7536529}.

Data backup is another category of mitigation techniques for ransomware attacks~\cite{cldsafe_an_efficient_file_backup}.
ShieldFS is a filesystem that automatically recovers files from backup copies, when the files are infected by ransomware~\cite{Continella:2016:SSR:2991079.2991110}.
However, data backup requires extra storage, and 
intelligent ransomware can erase backup copies~\cite{7924925,7886569}.

Mutual authentication techniques~\cite{Gotzfried:2014:MAT:2689660.2663348,4249828} are often 
used 
to protect disk drives.
The disk grants access
only if the host is authenticated by the protocol.
The host also accesses the disk only if the disk is authenticated.
The mutual authentication techniques are not intended to protect individual files
on the disk 
from malware.
Authentication entities are disks, not individual files.  
{\keyssd}

Encryption~\cite{6951337,7428036,1598131,6307756,6237012,Young:2015:DWE:2694344.2694387,6881505,7544407,6662530,7428037,7484726} is a popular technique 
for protecting data stored in 
disk drives.
The entire disk can be encrypted (full-disk encryption), or files are encrypted selectively.
Ransomware can still read files even if they are encrypted.
Ransomware cannot interpret the contents of the file, but can re-encrypt the file 
with its own encryption key.

Trusted Computing Module (TPM) is often used 
with authentication or encryption for key management~\cite{5689500}.
This is because 
keys need to be stored in a safe place. 
Since key management is orthogonal to our approach, TPM can also be 
used 
to manage keys for {\keyssd}.

%% file: conc.tex
\section{Concluding Remarks}
\label{sec:conc}
\vspace{-0.1in}

This paper proposes a fundamental countermeasure to ransomware attacks.
{\keyssd}, a disk drive using an access-control mechanism can be 
the last barrier of data breach.
Even if ransomware acquires an administrative privilege
and bypasses the access control mechanism in the file system, 
it cannot avoid the disk-level access control.
To support disk-level access control,
we slightly modify the Linux kernel to transfer the keys to access-control drives. 
Our extensive experimental results demonstrate that the performance overhead for {\keyssd} with {\statickeyftl} is negligible
and file data has been successfully protected from attacks by real ransomware samples.

%% file: ack.tex
\section{Acknowledgments }
\label{sec:ack}
\vspace{-0.1in}

We are grateful to Mr. Yung-woo Ko for proof-reading this paper. This work was supported by Samsung Semiconductor research grant.